\documentclass[aip,amsmath,amssymb,reprint,floatfix]{revtex4-1}

\usepackage[T1]{fontenc}
\usepackage{graphicx}
\usepackage{amssymb}
\usepackage{color}
\usepackage{times}
\usepackage{float}

\begin{document}

\newcommand{\bdt}[1]{\textcolor{black}{#1}}

\title{L\'evy noise-driven escape from \bdt{arctangent} potential wells}

\author{Karol Capa{\l}a}
\email{karol@th.if.uj.edu.pl}
\affiliation{Institute of Theoretical Physics, and Mark Kac Center for
Complex Systems Research, Jagiellonian University, ul. St. {\L}ojasiewicza 11,
30--348 Krak\'ow, Poland}
\author{Amin Padash}
\affiliation{Physics Department of Shahid Beheshti University, 19839-69411
Tehran, Iran}
\affiliation{Institute for Physics \& Astronomy, University of Potsdam,
14476 Potsdam-Golm, Germany}
\author{Aleksei V. Chechkin}
\affiliation{Institute for Physics \& Astronomy, University of Potsdam,
14476 Potsdam-Golm, Germany}
\affiliation{Akhiezer Institute for Theoretical Physics, 61108 Kharkov, Ukraine}
\author{Babak Shokri}
\affiliation{Physics Department of Shahid Beheshti University, 19839-69411
Tehran, Iran}
\affiliation{Laser and Plasma Research Institute, Shahid Beheshti University,
19839-69411 Tehran, Iran}
\author{Ralf Metzler}
\affiliation{Institute for Physics \& Astronomy, University of Potsdam,
14476 Potsdam-Golm, Germany}
\author{Bart{\l}omiej Dybiec}
\email{bartek@th.if.uj.edu.pl}
\affiliation{Institute of Theoretical Physics, and Mark Kac Center for
Complex Systems Research, Jagiellonian University, ul. St. {\L}ojasiewicza 11,
30--348 Krak\'ow, Poland}

\date{\today}

\begin{abstract}
The escape from a potential well is an archetypal problem in the study of stochastic dynamical systems, representing real-world situations from chemical reactions to leaving an established home range in movement ecology.
Concurrently, L{\'e}vy noise is a well-established approach to model systems characterized by statistical outliers and diverging higher-order moments, ranging from gene expression control to the movement patterns of animals and humans.
Here, we study the problem of L\'evy noise-driven escape from an almost rectangular, \bdt{arctangent} potential well restricted by two absorbing boundaries, \bdt{mostly under action of the Cauchy noise}.
We unveil analogies of the observed transient dynamics to the general properties of stationary states of L{\'e}vy processes in single-well potentials.
The first escape dynamics is shown to exhibit exponential tails.
We examine the dependence of the escape on the shape parameters, steepness and height, of the \bdt{arctangent} potential.
Finally, we explore in detail the behavior of the probability densities of the first-escape time and the last-hitting point.
\end{abstract}

\pacs{
05.40.Fb, % Random walks and Levy flights
05.10.Gg, % Stochastic analysis methods (Fokker--Planck, Langevin, etc.)
02.50.-r, % Probability theory, stochastic processes, and statistics
02.50.Ey, % Stochastic processes
}

\maketitle

%%%%%%%%%%%%%%%%%%%%%%%%%%%%%%%%%%%%%%%%%%%%%%%%%%%%%%%%%%%%%%%%%%%%%%%%

\textbf{The escape from a potential well is an archetypal process underlying many noise induced phenomena.
The escape protocol is sensitive to the noise type, therefore escape induced by equilibrium, thermal, Gaussian noise is very different to  escape
induced by non-equilibrium $\alpha$-stable white L\'evy noise driving.
In general, in the presence of Gaussian driving the height of the potential barrier determines the transition rate  while in the non-equilibrium regime the transition rate is determined by the barrier width.
As we show, in special situations, escape rates as well as first passage times are sensitive to both the width and height of the potential barrier --- as a particle may escape from the system not only via a single long jump by also following a sequence of short jumps.
The escape protocol affects not only the transition rates but also the time dependent probability densities.
Already for finite depths of a rectangular-like potential well, e.g., the \bdt{arctangent} potential well considered here, the part located within the potential well is similar to a stationary density recorded in a similar single-well potential.
At the same time, extra modes are placed outside the potential well.
Due to the presence of absorbing boundaries, the probability of finding a particle in the domain of motion decays exponentially over time, typical for Markovian diffusion in finite domains.
Finally, the escape scenario is also reflected in the last hitting point distribution.  
Escapes performed by a single-long jump are responsible for the emergence of a dominating peak at the initial position, while short jumps produce peaks in the vicinity of the absorbing boundaries.
}

\section{Introduction \label{sec:intro}}

After the theoretical description of thermal diffusion by Einstein, Sutherland, Smoluchowski, and Langevin \cite{einstein1905,sutherland1905,smoluchowski1906,langevin1908} in his seminal 1916 work Smoluchowski applied these insights to develop the theory of coagulation of two colloidal particles \cite{smoluchowski1916}.
His approach is still essential for many calculations of chemical reaction rates, and it is based on the first-hitting time of the two diffusing particles.
However, for a successful chemical reaction, in addition to the diffusion-limitation to find each other, in order to react the activation (Gibbs) energy barrier needs to be overcome, as originally proposed by Arrhenius \cite{arrhenius1889}.
Typically this requires the particles to collide multiple times before a successful reaction occurs \cite{collins1949,vonHippel1989facilitated}.
Considering this reaction-limited step on top of the diffusive search leads to a considerable further defocusing of the associated reaction-times \cite{Grebenkov2018,godec2016}, an important factor especially for reactions in the low-concentration limit, e.g., in gene expression \cite{Pulkkinen2013,Kolomeisky2011,Godec2017}.
Today, barrier crossing in chemical reactions (association and dissociation) is at the heart of reaction-rate theory \cite{hanggi1990}.
Originally worked out by Kramers in 1940 \cite{kramers1940} many additional facets of noise-driven barrier crossing have been explored.
Among the most prominent features, we mention stochastic resonance and resonant activation \cite{mcnamara1989,doering1992,gammaitoni1998}, breaking of detailed balance and thermal ratcheting \cite{magnasco1993,hanggi2009} underlying processes such as the motion of Brownian motors \cite{astumian2002}, and transition-state theory \cite{chandler1978,dellago2002}.

In the overdamped regime, the state of a classical particle is characterized by its position coordinate only.
In absence of noise the particle settles in any local minimum of the potential.  In this sense all minima of the potential are stable.
In contrast, in the presence of noise, e.g., thermal fluctuations, the relative stability of potential minima becomes modified, and typically deeper minima are more stable.
Moreover, noise may facilitate the escape from a given potential minimum.
Noise-driven escape is well studied \cite{hanggi1990} when the driving noise is white
and Gaussian \cite{gardiner2009,risken1996fokker}.
However the assumption of the Gaussianity of the noise might have to be relaxed
% as fluctuations don't necessarily need to be bounded.
as non-equilibrium fluctuations may often have large non-Gaussian outliers.
In fact, heavy-tailed distributions of fluctuations may still be white but approximated by L{\'e}vy-type, heavy-tailed noise densities generalizing Gaussian densities \cite{samorodnitsky1994}.
Conceptually, L{\'e}vy stable noise is utilized in many theoretical models \cite{jespersen1999,chechkin2000linear,chechkin2002b,ebeling2009microfields,ebeling2010convoluted,li2016levy,li2017transports,fogedby1998,fogedby1994}.
In fact, the generalized central limit theorem for identically distributed, independent variables with diverging variance gives rise to L{\'e}vy stable densities as limiting distributions \cite{feller1968,hughes1995}.
L{\'e}vy statistics may also emerge from deterministic nonlinear systems near a critical point \cite{Abe2020}. L{\'e}vy-type fluctuations have been observed in a wide variety of systems. These include fundamental physical systems such as heat transport anomalies \cite{cipriani2005}, transport in Lorentz-like gases \cite{barkai2000b} and weakly turbulent systems \cite{zaslavsky2005}, light propagation in disordered optical media \cite{barthelemy2008}, quantum optical systems \cite{katori1997}, and fluctuations in plasma devices \cite{Gonchar2003}. L{\'e}vy relocation statistics may emerge in dimensionally reduced systems, e.g., when excitations or molecules moving on long polymers may jump across shortcuts where the polymer loops back on itself \cite{sokolov1997,lomholt2005}.
In a biological context, we mention molecular-motor motion \cite{song2018neuronal,chen2015memoryless}, spreading of cancer cells \cite{Huda2018}, generalized models for gene transcription dynamics \cite{wang2018} and stability in gene regulatory networks \cite{luo2017}. In climate models L{\'e}vy noise represents extreme fluctuations \cite{ditlevsen1999}. An important field, in which L{\'e}vy stable relocation statistics have been widely explored, is macroscopic movement.
These systems include individual albatross birds and sea predators \cite{humphries2012,humphries2010,barthelemy2008,viswanathan2011physics}, human hunter-gatherer foraging \cite{raichlen2014}, pedestrian movement \cite{murakami2019}, modern-day human movement dynamics \cite{brockmann2006}, but also optimal robotic search \cite{fioriti2015}.
Further applications of non-local, L{\'e}vy-type search are found in computer algorithms such as simulated annealing \cite{pavlyukevich2007b}. Recently, evidence for L{\'e}vy-type statistics was reported in the Covid-19 pandemic propagation \cite{gross2020spatio}.

Other prominent properties of systems driven by $\alpha$-stable noises are related to stationary states in single-well potentials.
For Gaussian white noise in such potential, stationary states are of the Boltzmann--Gibbs type, i.e., they are unimodal.  In the non-equilibrium regime, i.e., under action of $\alpha$-stable noises, if stationary states exist they are not of the Boltzmann--Gibbs type \cite{Gonchar2002,chechkin2002}.
Particularly, they can be multimodal \cite{chechkin2003,chechkin2004,dubkov2007,dubkov2008,capala2019multimodal}.
Multimodality of stationary states emerges due to the competition between deterministic and random forces. The deterministic force is the restoring force, while the random force is responsible for long excursions.
If a particle cannot return to the origin before the next noise-induced excursion, for $V(x)=|x|^{\nu}$ with $\nu>2$, the stationary state is multimodal with modes located in the vicinity of maxima of the potential curvature \cite{chechkin2003,chechkin2004,capala2019multimodal}.
Moreover, the scenario of emergence of bimodal stationary states has unexpected properties, because the transient densities  between initial delta peak and final bimodal distribution can be trimodal \cite{chechkin2004}.
The trimodal transient density appears, because the initial peak disappears slower than the outer
modes emerge.

L{\'e}vy noise-driven barrier escape, the central topic of this study, in the above-mentioned systems is relevant, inter alia, for the departure from locally stable climate states as outlined in \cite{ditlevsen1999}, or as a proxy for the escape of animals from their home range.
Studying the barrier crossing dynamics may shed new light on confinement of plasmas, or the transport of heat across insulating layers.
Similarly it might help to understand how in genetic systems seemingly stable states may be left.
The barrier crossing dynamics of a stochastic system driven by L{\'e}vy-stable white noise is significantly different from the scenario under Gaussian noise \cite{ditlevsen1999,imkeller2006b,chechkin2005,chechkin2007,imkeller2010hierarchy,dubkov2008,dubkov2009,dubkov2010}.
Thus, in the Gaussian case the continuous particle trajectories force the particle to actually surmount the potential barrier, such that the mean escape time depends exponentially on the depth of the potential well \cite{kramers1940,hanggi1990}.
In contrast, under L{\'e}vy-stable noise the trajectories exhibit long jumps \cite{koren2007,dybiec2016jpa}, and the particle may simply jump across the barrier without actually reaching the top of the potential barrier \cite{ditlevsen1999}.
Due to the possibility of anomalously long jumps the escape time becomes sensitive to the \emph{width\/} of the potential barrier \cite{ditlevsen1999,imkeller2006,imkeller2010hierarchy,bier2018}, and the dependence of the mean escape time is, approximately, inversely proportional to the noise strength \cite{chechkin2007}.

Here, we explore several aspects of the barrier crossing dynamics of stochastic processes driven by L{\'e}vy-stable noise, using a specific \bdt{arctangent} potential well in a finite interval delimited by two absorbing boundaries.
Putting these boundaries a distance from the edges of the potential well, we are able to follow a rich time evolution of the probability densities and uncover interesting properties of the escape dynamics.
In particular, we disclose the dependence of the escape times on both the width and depth of the potential.
Apart from the escape times we also examine the last-hitting point distribution.
For the probability densities we find multimodal states reflecting the competition between long jumps and confinement by the potential.
This relates the examination of the noise-induced escape with general properties of stationary states in single-well potentials.
Our results from extensive Monte Carlo simulations are compared with numerical solutions of the space-fractional Smoluchowski-Fokker-Planck equation.

The paper is structured as follows: in Section \ref{sec:model} we formulate the model and review some properties of L{\'e}vy noise-driven motion in confining potentials relevant for the study of the time-dependent probability densities.
In Section \ref{sec:results} we then present our main results corresponding the Cauchy ($\alpha=1$) noise for the pre-asymptotic system (Sec.~\ref{sec:pdf}), the last-hitting point behavior (Sec.~\ref{sec:lhp}), and the escape time statistics (Sec.~\ref{sec:fpt}).
Section~\ref{sec:neq1} shows similarities and differences between Cauchy ($\alpha=1$) and general $\alpha$-stable driving.
Finally, Section~\ref{sec:summary} concludes the paper.

\section{Formulation of the model \label{sec:model}}

The overdamped Langevin equation
\begin{equation}
     \dot{x}(t) = -V'(x) + \xi (t).
    \label{eq:langevin}
\end{equation}
is the typical starting point for the description of a large variety of stochastic dynamical systems \cite{gardiner2009,coffey2012langevin}.
It governs motion of a test particle under the combined action of a deterministic potential force $-V'(x)$ and a random force $\xi(t)$.
The stochastic force approximates the complex interactions of the test particle with its environment.
Here, we assume that noise $\xi(t)$ is of $\alpha$-stable type, i.e., it is the formal time derivative of the symmetric $\alpha$-stable process $L(t)$ \cite{janicki1994,dubkov2008} whose increments $\Delta L=L(t+\Delta t)-L(t)$ are independent and identically distributed according to the $\alpha$-stable density.
We restrict ourselves to symmetric $\alpha$-stable noises only.  Symmetric $\alpha$-stable distributions are unimodal densities with the characteristic function \cite{samorodnitsky1994,janicki1994}
\begin{equation}
\varphi(k)=\langle e^{ik\Delta L} \rangle = \exp\left[ -\Delta t \sigma^{\alpha}|k|^{\alpha} \right].
    \label{eq:levycf}
\end{equation}
The stability index $\alpha$ ($0<\alpha \leqslant 2$) determines the tail of the distribution, which for $\alpha<2$ is of power-law type $\varphi(x) \simeq |x|^{-(\alpha+1)}$.
The positive parameter $\sigma$ is the scale parameter, i.e., it controls the width of the distribution, typically defined by an interquantile width or by fractional moments, as variance of $\alpha$-stable variables with $\alpha<2$ diverges.
\bdt{In the numerical studies presented below the scale parameter is set to unity, i.e., $\sigma=1$.}
\bdt{Moreover, we mainly focus on the case of Cauchy noise, i.e., $\alpha$-stable noise with $\alpha=1$.
Nevertheless, for illustration we show some results corresponding to other values of the stability index $\alpha$, including $\alpha=2$, which corresponds to the Gaussian white noise driving.
}

\begin{figure}[H]
    \centering
    \includegraphics[width=0.80\columnwidth]{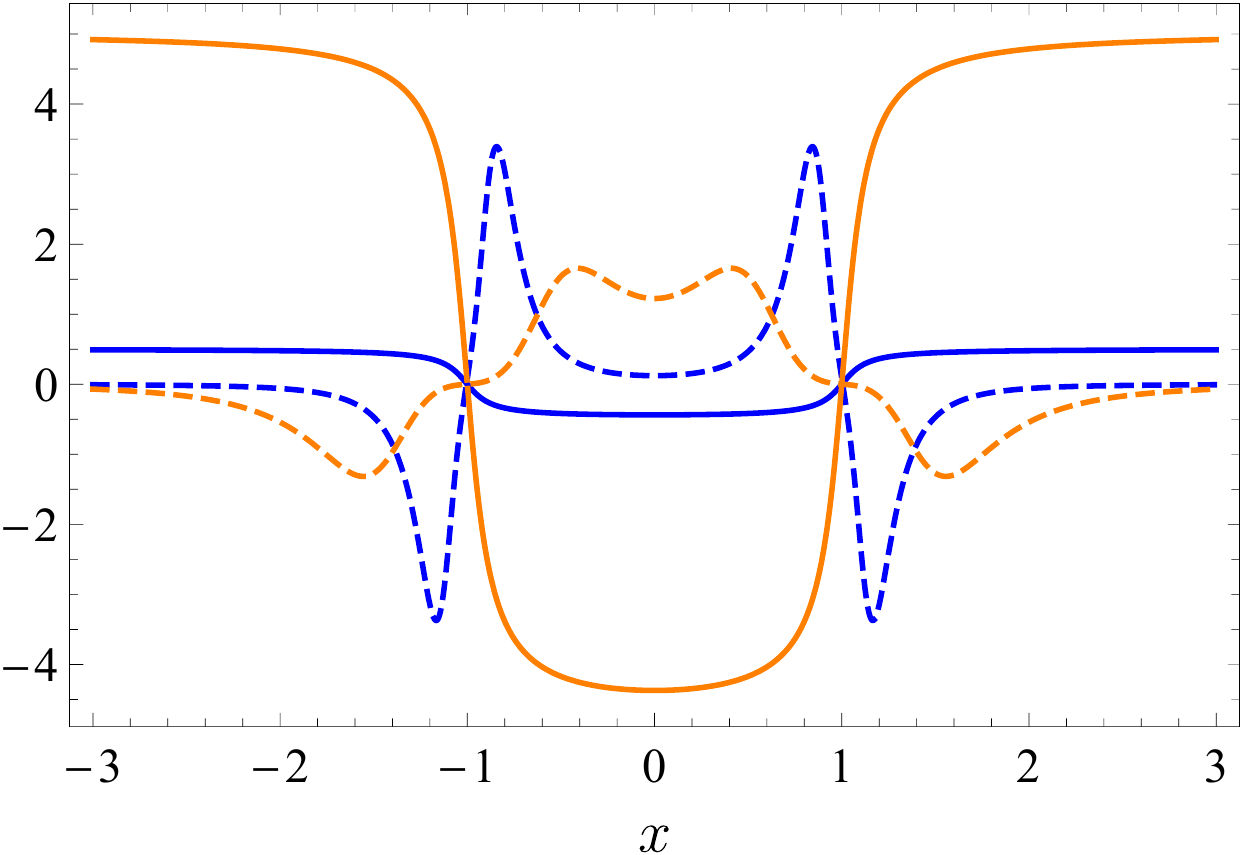}\\
    \includegraphics[width=0.80\columnwidth]{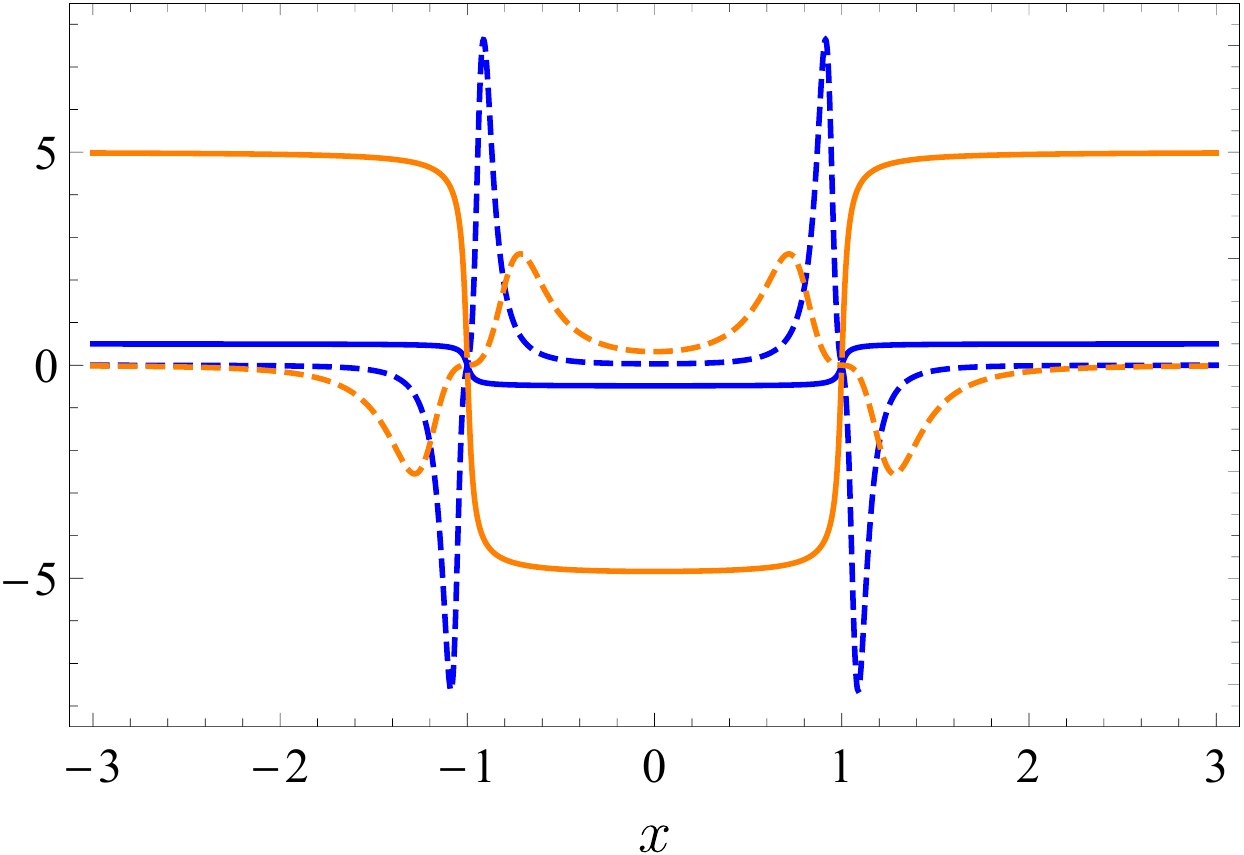}
    \caption{Potential $V(x)$ (solid lines) and potential curvature $\kappa(x)$ (dashed lines) for $n=5$ (top panel) and $n=20$ (bottom panel).
    Blue lines correspond to $h=1$, orange lines to $h=10$.}
    \label{fig:potentials}
\end{figure}

Within the current study we assume that the random walker moves within a bounded domain, restricted by two absorbing boundaries, in the particular external, arctan, potential
\begin{equation}
\label{eq:potential}
V(x)=\frac{h}{\pi} \arctan{\left( nx^2-n\right)}.
\end{equation}
The shape of the potential Eq.~(\ref{eq:potential}) is determined by  the two parameters $n$ and $h$.
Parameter $n$ controls the steepness of the potential --- the larger the value of $n$ the steeper the potential is around $|x|=1$, and, in the limit $n\to \infty$, it becomes the rectangular potential well.
Parameter $h$ characterizes the depth of the potential well.
In other words, changes in $n$ affect the width of the potential barrier, within $x \approx \pm 1$ beyond which the potential rapidly grows.
This rapid growth of the potential is associated with a significant value of the deterministic force.
At the same time, changes in $h$ affect the barrier height.
\bdt{For large $n$ and $h\to\infty$ the potential given by Eq.~(\ref{eq:potential}) merges into an infinite rectangular potential well, which is frequently applied in quantum mechanics \cite{liboff2003introductory}, the study of hard sphere systems \cite{alder1957phase}, the Percus--Yevick approximation \cite{baxter1968percus} or the study of stationary states under L\'evy noises \cite{denisov2008}, to name a few.}
\bdt{For finite $h$ the potential can be used to approximate a finite rectangular potential barrier.}
Therefore, the potential given by Eq.~(\ref{eq:potential}) is especially suitable to study various hypotheses regarding the escape protocol, \bdt{as it allows for easy control of the depth and width of the potential barrier}.
Exemplary potentials of the form given by Eq.~(\ref{eq:potential}) are depicted in Fig.~\ref{fig:potentials} along with corresponding curvatures $\kappa(x)$
\begin{equation}
\kappa(x)=\frac{V''(x)}{\left[1+V'(x)^2 \right]^{3/2}}.
    \label{eq:curvature}
\end{equation}
The curvature $\kappa(x)$ plays an important role in determining the shape of stationary states in single-well potentials \cite{chechkin2003,capala2019multimodal}, as modal values of the stationary densities can be attributed to extremes of curvatures.
Since the motion is restricted to a finite domain by the two absorbing boundaries there are no stationary states in this system.
Nevertheless, the time dependent densities can still be multimodal.
As it will be shown below, the location of the modes can be still attributed to maxima of the potential curvature, see Eq.~(\ref{eq:curvature}).

The Langevin equation~(\ref{eq:langevin}) takes the following discretized \cite{janicki1994,janicki1996} form
\begin{equation}
    x(t+\Delta t) = x(t) - V'(x) \Delta t + \xi_t \times (\Delta t)^{1/\alpha},
    \label{eq:discretization}
\end{equation}
where $\xi_t$ represents a sequence of independent identically distributed symmetric $\alpha$-stable random variables \cite{chambers1976,weron1995,weron1996}, i.e., their characteristic function is given by Eq.~(\ref{eq:levycf}) with $\Delta t=1$ (as $\Delta t$ is explicitly present in Eq.~(\ref{eq:discretization})).
The particle starts its motion in the center of the potential well, i.e., $x(0)=0$, while the absorbing
boundaries are located at $x=\pm 3$.
Due to the presence of the absorbing boundaries the motion is continued as long as $|x|<3$.
\bdt{Such a location of the absorbing boundaries is motivated by two criteria: (i) it cannot be too far from the origin because the MFPT grows with increasing interval half-width, see Eq.~(\ref{eq:MFPTfree}), and (ii) it needs to be far enough to assure that there is a flat part of the potential. 
The positions of the absorbing boundaries at $\pm3$ assure  that for any reasonably large $n$ there is a flat part of the potential outside the potential well, i.e., the absorbing boundaries are outside the potential well and the average escape time is not overly large. 
Under such conditions, the obtained results do not change qualitatively with the change in the location of potential barriers.}
Every time a particle crosses the absorbing boundary, it is immediately removed from the system.
Consequently, the amount of particles present in the system decays over time.
Asymptotically, for $t\to\infty$ all particles escape from the potential well.
As we already mentioned in the Introduction, taking into account both dynamics inside the \bdt{arctangent} potential well and on the flat part of the potential allows us to unveil a rich and interesting time evolution determined by the mixture of these two escape problems.

Equation~(\ref{eq:langevin}) describes the system's time evolution on a single trajectory level.
From ensemble of trajectories $x(t)$, it is possible to obtain the macroscopic evolution of the probability density $P(x,t|x_0,0)=\langle \delta(x-x(t)) \rangle$.
The evolution of the probability density $P(x,t|x_0,0)$ is provided by the space-fractional Smoluchowski-Fokker-Planck \cite{metzler1999,schertzer2001,yanovsky2000} equation
\begin{equation}
    \frac{\partial P}{ \partial t} = \frac{\partial}{\partial x} \left[ V'(x)P \right] +\sigma^\alpha \frac{\partial^\alpha P}{\partial |x|^\alpha}.
    \label{eq:ffp}
\end{equation}
The space-fractional derivative $\partial^\alpha P/\partial |x|^\alpha$ of the Riesz-Weyl type \cite{podlubny1999,samko1993} can be defined by its Fourier transform
\begin{equation}
    \mathcal{F}\left[\frac{\partial^\alpha P}{\partial |x|^\alpha}\right]= -|k|^\alpha \mathcal{F}(P).
    \end{equation}
For details of the numerical method for solving Eq.~(\ref{eq:ffp}) we refer the reader to \cite[Appendix~B]{padash2019first}.

A particle moving in the potential (\ref{eq:potential}) restricted by two absorbing boundaries will surely leave the interval $[-3,3]$.
For such a system, it is possible to calculate the survival probability
\begin{equation}
    S(t)=\int_{|x|<3}P(x,t|x_0,0)dx,
    \label{eq:survival}
\end{equation}
which gives the fraction of particles that at time $t$ are still in the system,
i.e., in the interval $[-3,3]$.
Moreover, using the survival probability it is possible to define quasi-stationary state \cite{siler2018,ornigotti2018,ryabov2019}
\begin{equation}
    Q(x)= \lim_{t\to\infty} \frac{P(x,t|x_0,0)}{S(t)},
    \label{eq:quastiStationaryStates}
\end{equation}
which can be conveniently used as the auxiliary quantity in examination of some time dependent probability densities.

\section{Results for the Cauchy case $\alpha=1$ \label{sec:results}}

\bdt{The problem of escape from the arctangent potential well we consider here shares some features of solvable models driven by L\'evy noises: free motion \cite{samorodnitsky1994}, escape of a free particle from a finite interval restricted by two absorbing boundaries \cite{blumenthal1961,getoor1961} and stationary states in an infinite rectangular potential well \cite{denisov2008} or in single-well potentials of $V(x)\propto x^{2m}$ type under Cauchy noise\cite{dubkov2007}.
Nevertheless, the model is studied numerically, as it does not fully belong to any solvable case and it does not allow for the use of approximations developed in Refs.~\onlinecite{ditlevsen1999,bier2018, imkeller2006,imkeller2006b}, because, as it is demonstrated below, the mean first passage time is sensitive both to the width and height of the potential barrier.
}
We start by studying the detailed properties of the time dependent densities for the generic case $\alpha=1$ (Sec.~\ref{sec:pdf}).
Next, we switch to the escape kinetics by exploring the last hitting point distributions (Sec.~\ref{sec:lhp}) and the first passage times along with the \bdt{mean first passage time (MFPT)} (Sec.~\ref{sec:fpt}).
As the driving noise in Sections \ref{sec:pdf} --- \ref{sec:fpt} we use Cauchy noise, i.e.,
$\alpha$-stable noise with $\alpha=1$. The details of the escape kinetics
for $\alpha\neq 1$ in our setup require separate study and will be discussed
in Sec.~\ref{sec:neq1}.
\bdt{Within the following studies we set the scale parameter $\sigma$ to unity i.e. $\sigma=1$.}

\subsection{Time dependent densities \label{sec:pdf}}

\begin{figure}[h!]
\begin{tabular}{c c}
\includegraphics[width=0.45\columnwidth]{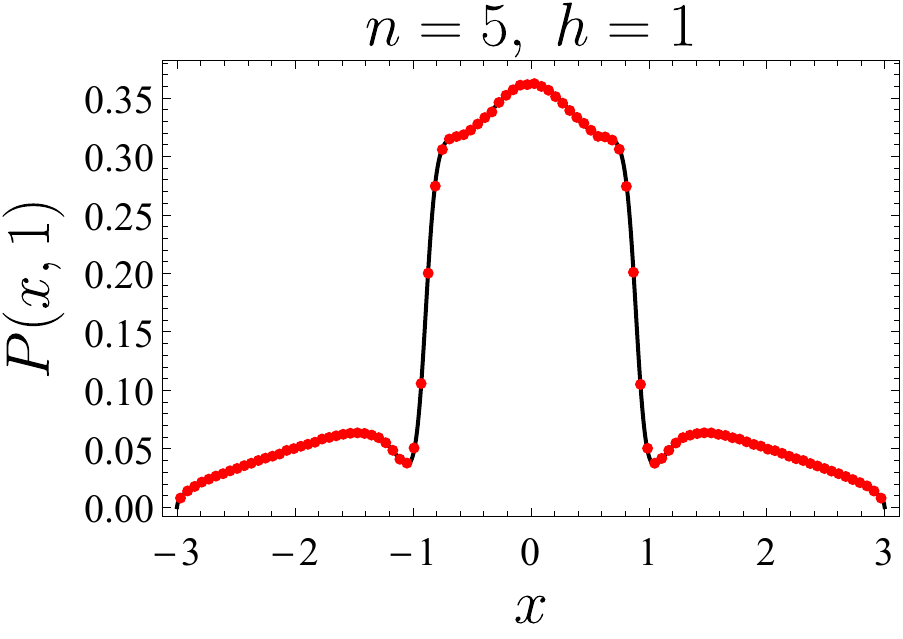}&
\includegraphics[width=0.45\columnwidth]{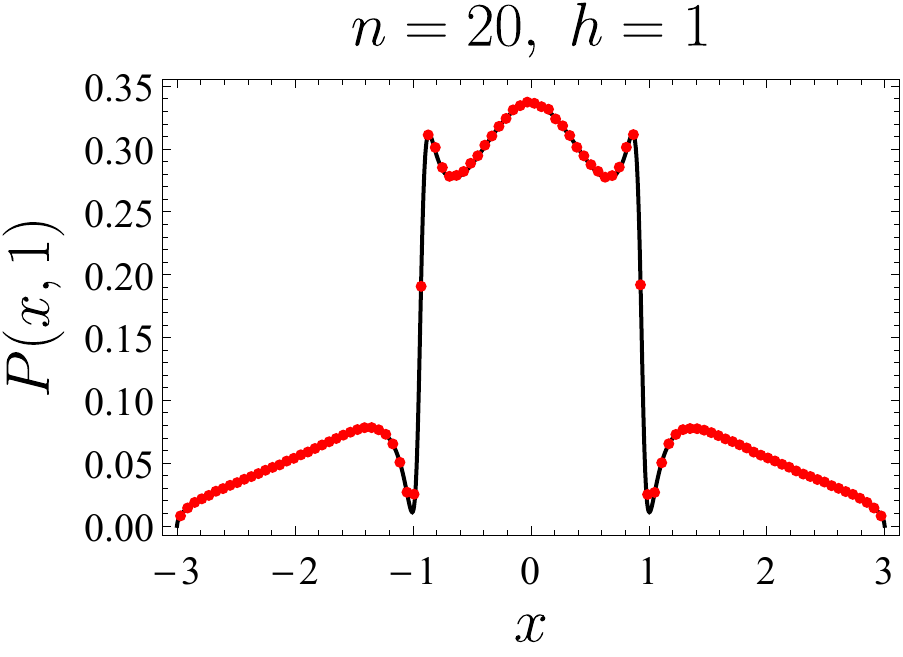}\\
\includegraphics[width=0.45\columnwidth]{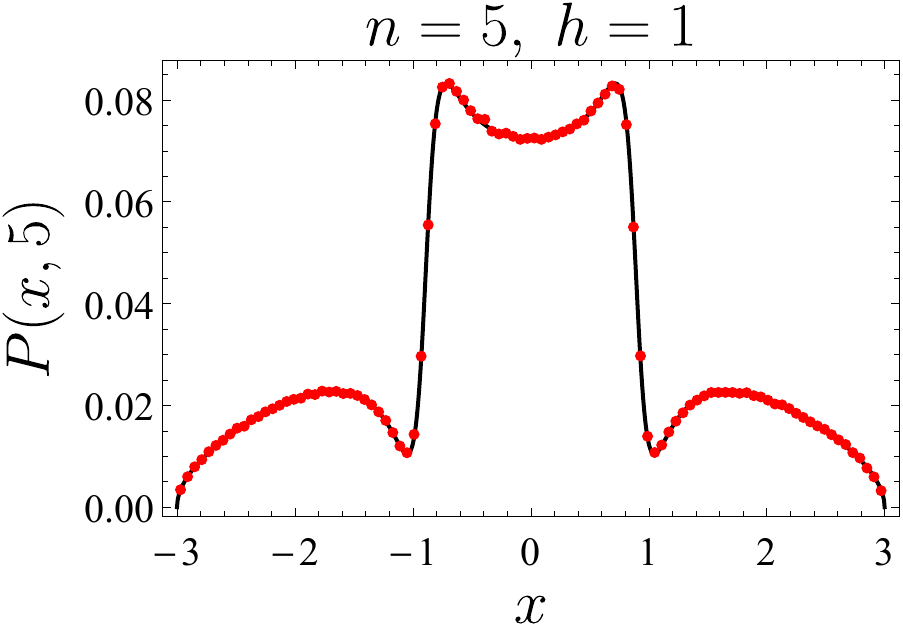}&
\includegraphics[width=0.45\columnwidth]{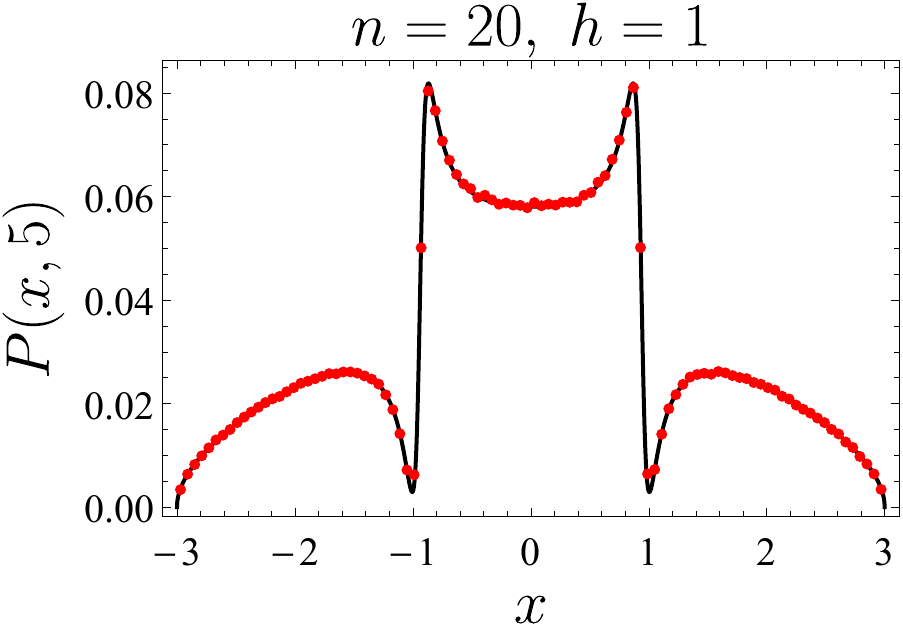}\\
\end{tabular}
\caption{Time dependent probability densities \bdt{$P(x,t|x_0=0,0) \equiv P(x,t)$} corresponding to various parameters $n,h$ characterizing the potential~(\ref{eq:potential}) at various times $t$.
In the plots the red points denotes results obtained by stochastic simulation of the Langevin equation~(\ref{eq:discretization}), while the black solid lines represent the numerical solution of the corresponding space-fractional Smoluchowski-Fokker-Planck equation~(\ref{eq:ffp}).
The panels show the time dependent densities at times  $t=1$ (top) and $t=5$ (bottom).
The columns correspond to $n=5$ (left) and $n=20$ (right).}
\label{fig:pdfsh1}
\end{figure}

\begin{figure}[h!]
\begin{tabular}{c c}
\includegraphics[width=0.45\columnwidth]{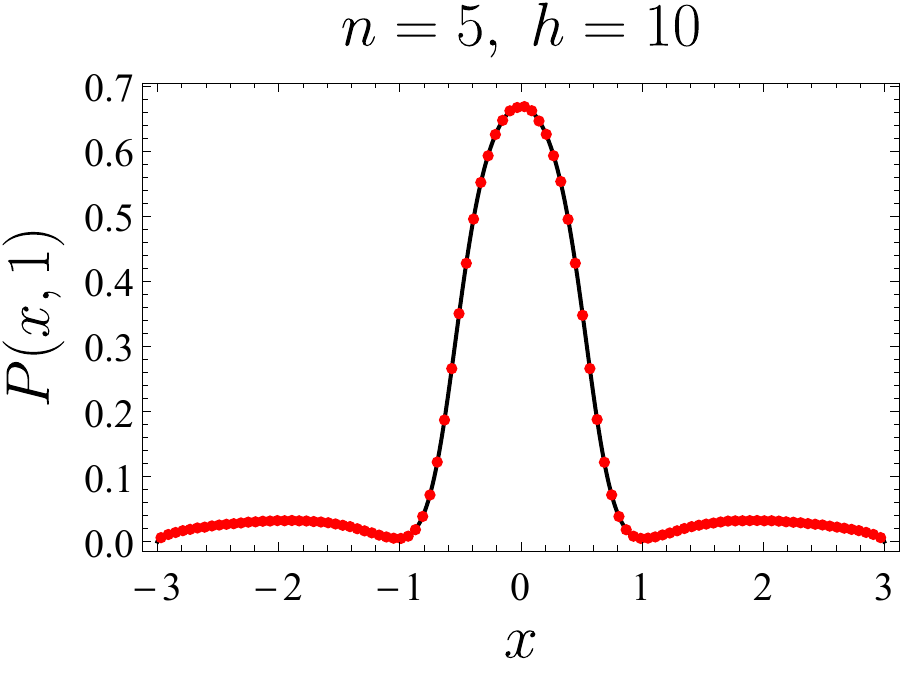}&
\includegraphics[width=0.45\columnwidth]{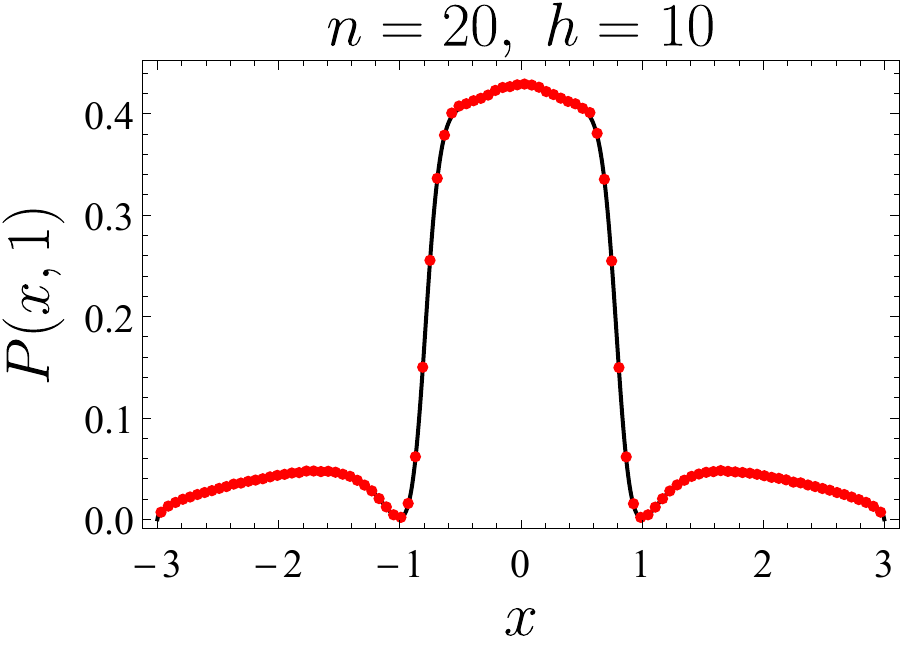}\\
\includegraphics[width=0.45\columnwidth]{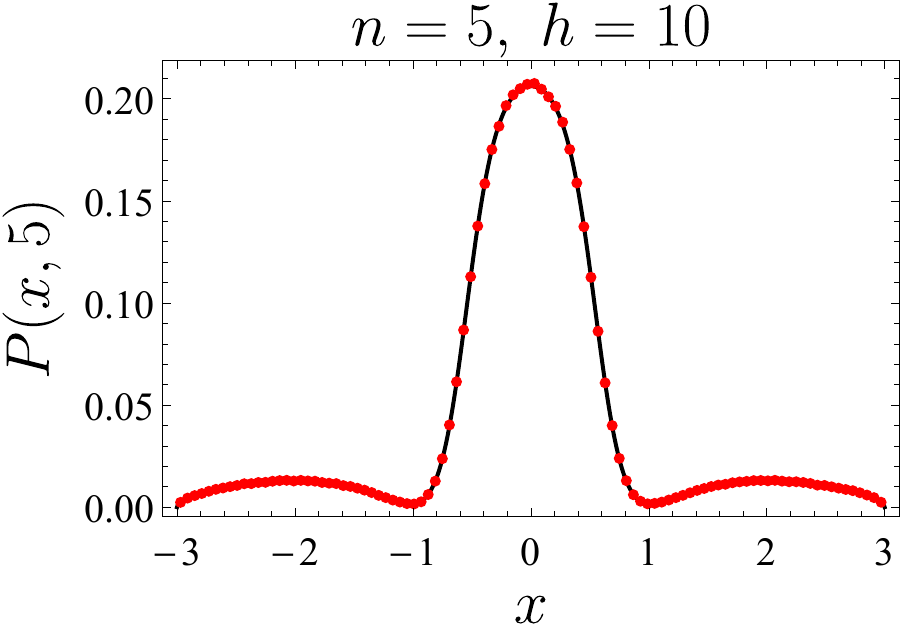}&
\includegraphics[width=0.45\columnwidth]{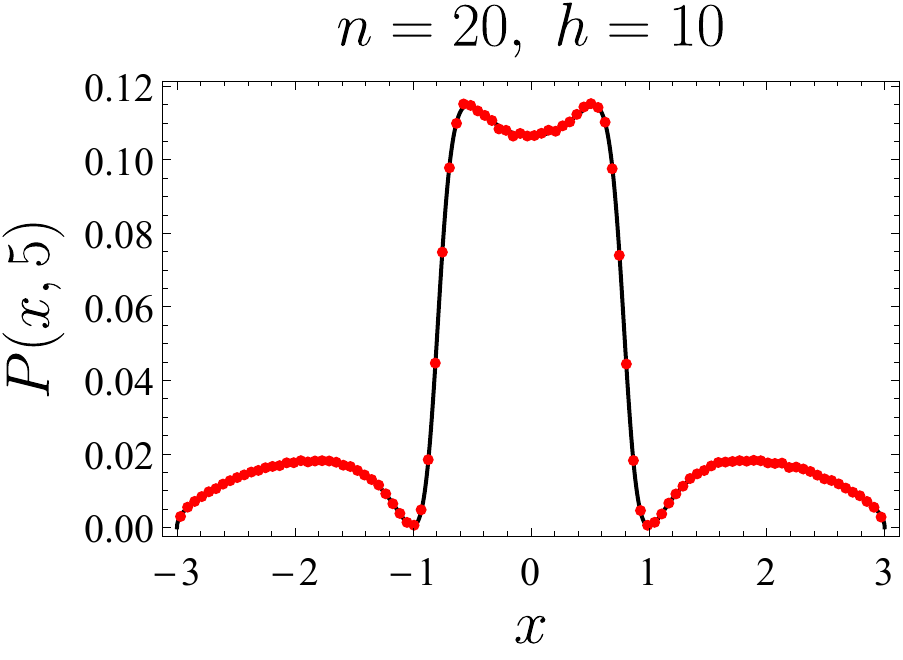}
\end{tabular}
\caption{The same as in Fig.~\ref{fig:pdfsh1} for $h=10$.}
\label{fig:pdfsh10}
\end{figure}

Time dependent densities $P(x,t|x_0=0,0)$ have been constructed numerically by ensemble averaging of trajectories $x(t)$ generated by Eq.~(\ref{eq:discretization}) and by numerical methods for the fractional Smoluchowski-Fokker-Planck equation \cite[Appendix~B]{padash2019first}.
Sample time-dependent densities are depicted in Figs.~\ref{fig:pdfsh1} -- \ref{fig:pdfsh10}.

The time dependent densities are constructed for the symmetric initial condition, $x(0)=0$, which corresponds to $P(x,0|0,0)=\delta(x)$.
Therefore, the shape of densities is determined by the interplay of three processes: the decay of the initial peak, the emergence of outer peaks within the potential well and the ultimate absorption of particles.
The time dependent densities consist of two parts.
The first part is located within the potential well, with $x\in (-1,1)$, while the second, outer part corresponds to $1<|x|<3$.
The evolution of the inner part ($|x|<1$) is related to the scenario of reaching stationary states in single-well potentials \cite{metzler2004,chechkin2004}, while the outer part ($|x|>1$) is determined by the jump length distribution.
As the domain of motion is restricted by two absorbing boundaries, asymptotically all particles escape from the domain of motion.
Consequently, the number of particles decays over time.  Nevertheless, for $t\geqslant 2$ the time dependent densities do not perceivable change their shape.
This phenomena reflects the effect that quasi-stationary state $Q(x)$ has been reached, see Eq.~(\ref{eq:quastiStationaryStates}).

In the scenario of emergence of bimodal stationary densities in single-well potentials of $|x|^\nu$ ($\nu>2$) two regimes are observed \cite{chechkin2004}.
For $\nu>4$ the initial peak crosses over into a final bimodal state via a transient trimodal state, while for $2< \nu \leqslant 4$ there is direct crossover from uni- to bimodal state \cite{chechkin2004}.
Analogous scenarios are recorded for the inner part of the time dependent densities, for $|x|<1$, see Fig.~\ref{fig:pdfsh1} (transient trimodal state) and Fig.~\ref{fig:pdfsh10} (transient bimodal state).
With increasing $n$, the outer peaks move further towards the barrier \cite{denisov2008,ciesla2019multimodal}, i.e., they approach $x=\pm 1$.
Moreover, peaks located outside the potential well, on the flat part of the potential profile, are amplified because with the increasing $n$ (with increasing steepness of the potential barrier) the potential curvature is also increased.

The analysis of the potential curvature $\kappa(x)$ suggests that for finite $n$ it is possible to select  $h$ such that the inner ($|x|<1$) part of the time dependent densities will be unimodal.
Indeed, such a behavior is seen in Fig.~\ref{fig:pdfsh10} which presents the same results as Fig.~\ref{fig:pdfsh1} but for $h=10$ instead of $h=1$.
With increasing $h$ we observe a weakening of the modes of time dependent densities at $|x|\approx 1$.
Finally, for $h$ large enough the inner part of the time dependent densities becomes unimodal.
Importantly, we observe deviations from the typical, $\nu>4$, crossover scenario, mentioned above, because transient states are invariably unimodal: we do not see a trimodal
transient state.
Nevertheless, for a fixed $h$ the bimodality of the inner part of the probability density can be reintroduced, see bottom left and bottom right panels of Fig.~\ref{fig:pdfsh10} in which $n$ is increased from $n=5$ to $n=20$.
Therefore, we conclude that, contrary to stationary states in single-well potentials, the modality of the inner ($|x|<1$) part  of time dependent densities is determined both by the steepness of the potential well ($n$) and height of the potential barrier ($h$).

The local maxima of $P(x,t|x_0=0,0)$ at $|x|>1$ are produced by those particles which managed to jump out of the potential well and landed on the flat part of the potential.
The deterministic force and the random force produced by the central part of $\alpha$-stable density produce minima of the time dependent densities at $x=\pm 1$, as the deterministic force with a little help of random force move particles which landed at $|x| \approx \pm 1 $ back to the potential well.
With increasing $n$ the barrier width decreases, locating the minima and outer maxima closer to the barriers at $x=\pm 1$.
Moreover, minima become deeper and the maxima higher.

The shape of the time dependent densities for $|x|>1$ can be explained by the properties of $\alpha$-stable densities, which induce random jumps.
On the one hand, the central part $x\approx 0$ of the jump length distribution is responsible for the penetration of the system. \textcolor{black}{Subsequently, particles which penetrate the system are responsible for tails of the survival probability as they are absorbed after a longer time.}
On the other hand, the tails of the jump length distribution allow a particle to leave the domain via a single long jump.
The intermediate part of the jumps length distribution produces local maxima of $P(x,t|x_0=0,0)$ at $1<|x|<3$.
Due to the monotonous decay of $\alpha$-stable densities, in general, time dependent densities on the flat part of the potential decay with increasing $|x|$.
Local minima of the time dependent densities are placed close to the barrier, i.e., at $x\approx \pm 1$, because of the finite width of the barrier and the central part of jump length distribution which move some of particles back to the potential well.

\begin{figure}[H]
\centering
\includegraphics[width=0.23\textwidth]{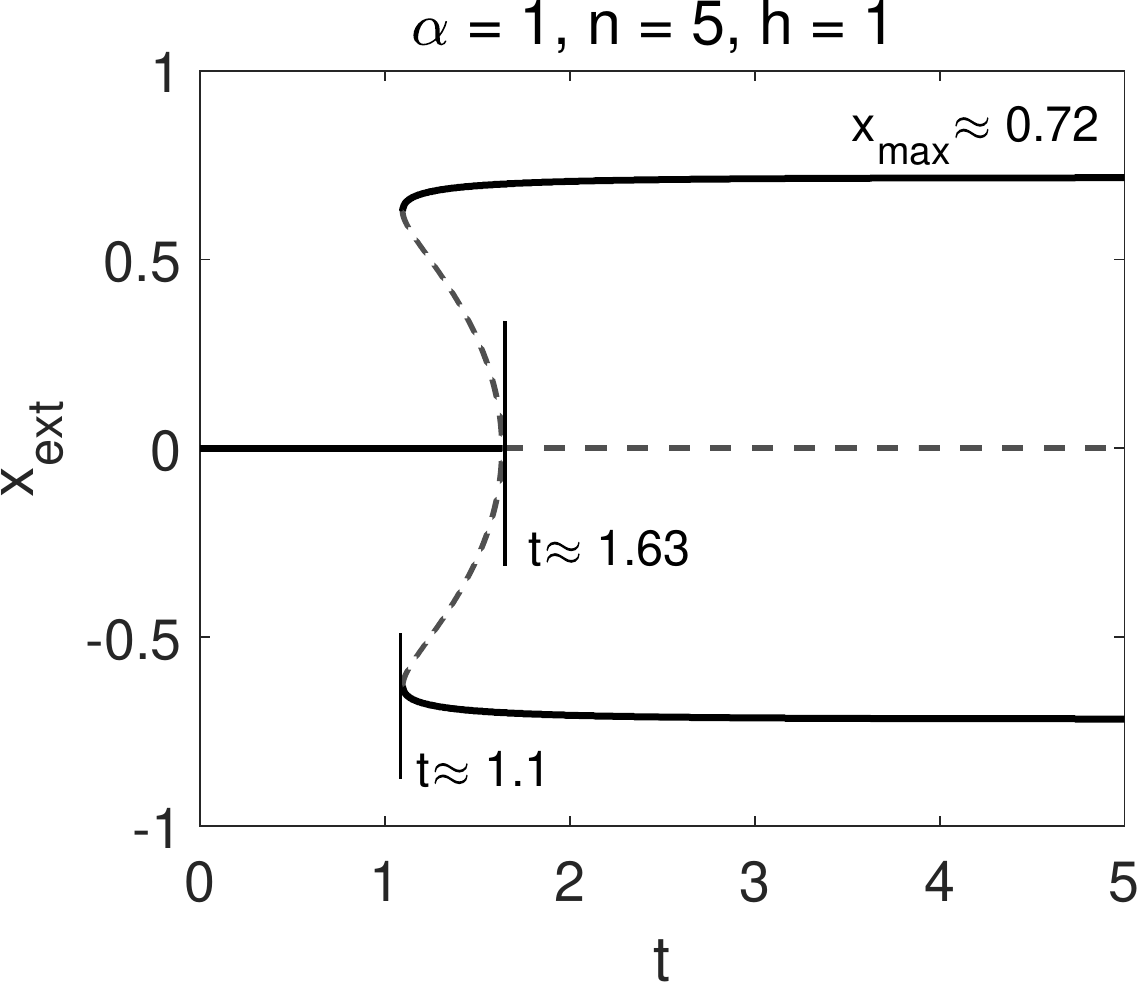}
\includegraphics[width=0.23\textwidth]{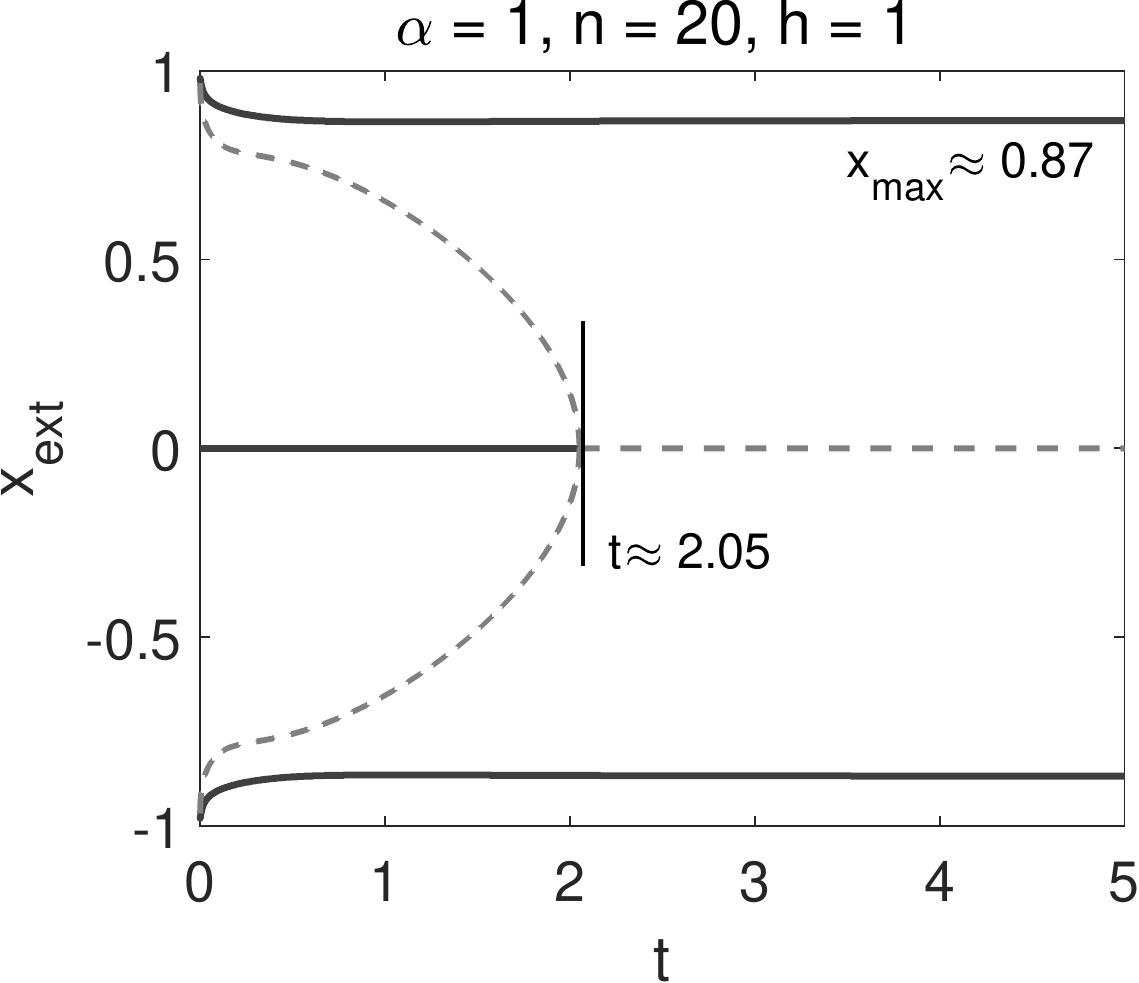}\\
\includegraphics[width=0.23\textwidth]{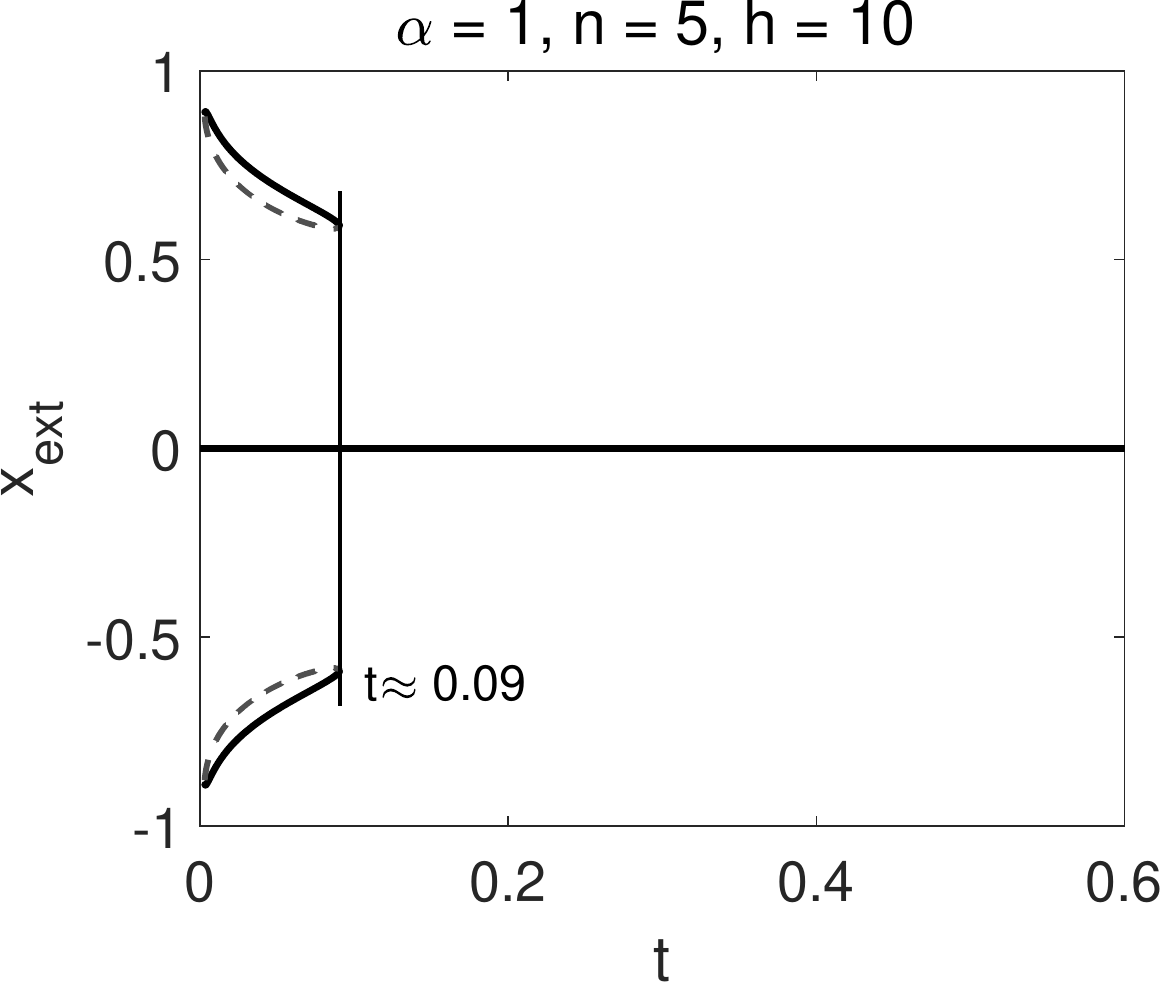}
\includegraphics[width=0.23\textwidth]{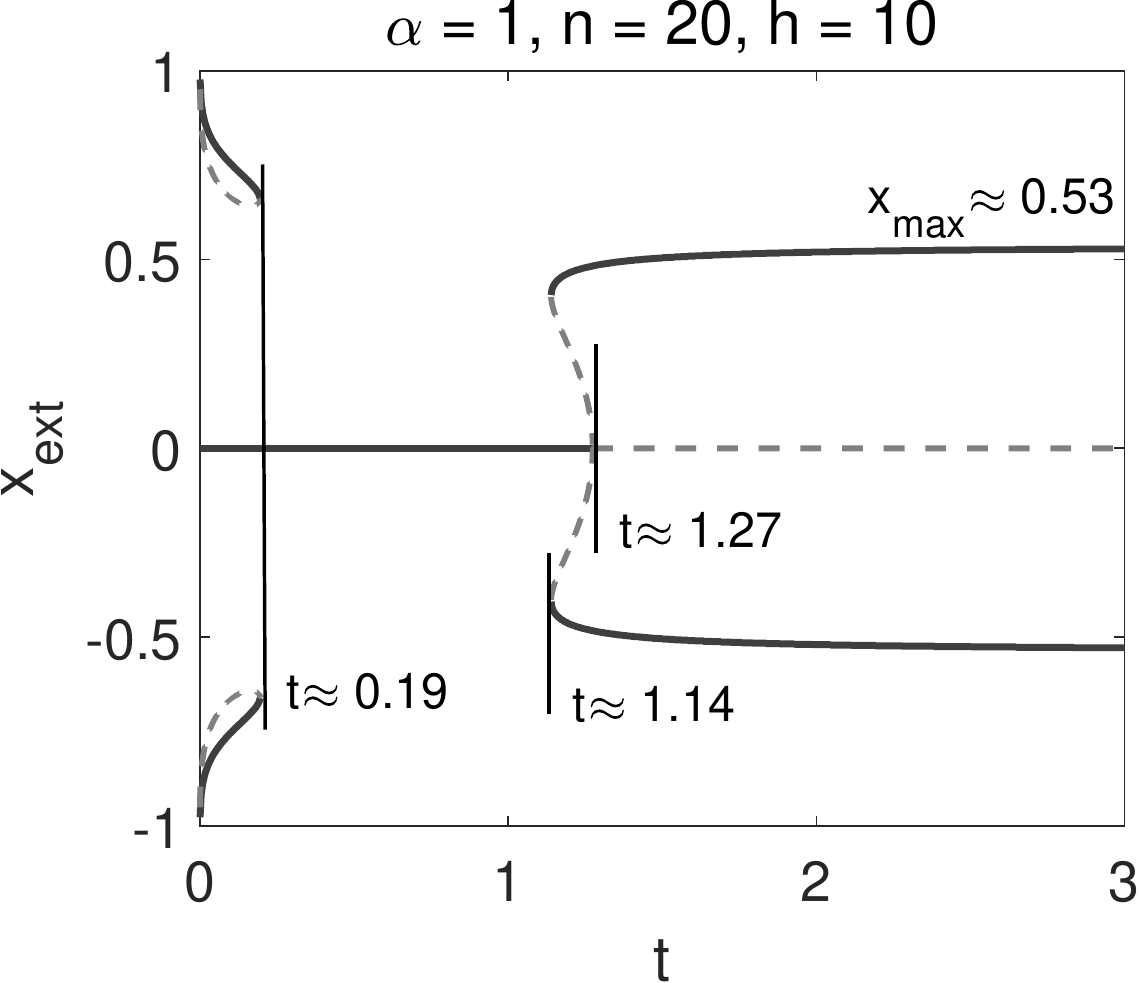}
\caption{Bifurcation diagrams for stability index $\alpha=1$ corresponding to $|x|<1$ parts of Figs.~\ref{fig:pdfsh1} and \ref{fig:pdfsh10}. Solid lines show the positions $x_{\max}$ of the maxima (global and local), the dashed lines indicate positions of the minima (global and local).
The panels show the bifurcation times for $h=1$ (top) and $h=10$ (bottom).
Columns correspond to $n=5$ (left) and $n=20$ (right).
The thin vertical lines indicate the bifurcation times.
The results are obtained by numerical solution of the space-fractional Smoluchowski-Fokker-Planck equation (\ref{eq:ffp}) with time step $\Delta t=0.0001$ and space increment $\Delta x=0.001$.}
\label{fig:bdiagram}

\end{figure}

Finally, Fig.~\ref{fig:bdiagram} shows the bifurcation diagrams corresponding to inner ($|x|<1$) parts of time dependent densities from Figs.~\ref{fig:pdfsh1} and~\ref{fig:pdfsh10}.
They display the locations of the maxima (solid lines) and the minima (dashed lines) of the time dependent densities.
From Fig.~\ref{fig:bdiagram} it is clearly visible that the time dependent densities can attain various multimodal states.
Typically, after sufficiently long time, they are bimodal, but they can also be unimodal, see the bottom left panel of Fig.~\ref{fig:bdiagram}.
Moreover, the phenomenology of the transient probability densities can be very different: depending on parameters characterizing the potential, can change their modality from
$1\to 3\to 2$ (top left),
$3 \to 2$ (top right),
$3 \to 1$ (bottom left)
and
$3 \to 1 \to 2$ (bottom right) modes.
Additionally, in Fig.~\ref{fig:bdiagram} crossover times are included.

\subsection{Last hitting point density\label{sec:lhp}}

\begin{figure}[H]
\includegraphics[width=0.80\columnwidth]{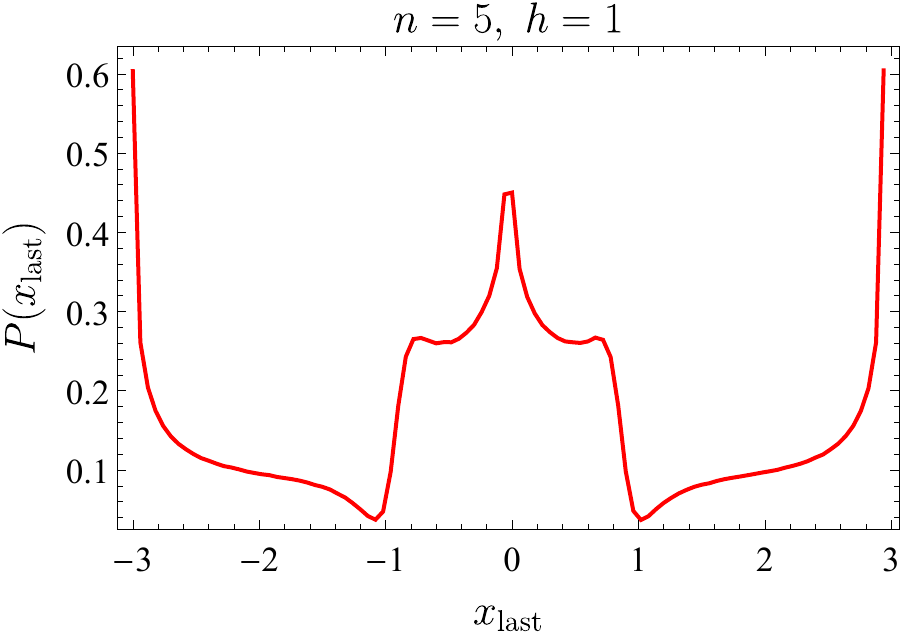}\\
\includegraphics[width=0.80\columnwidth]{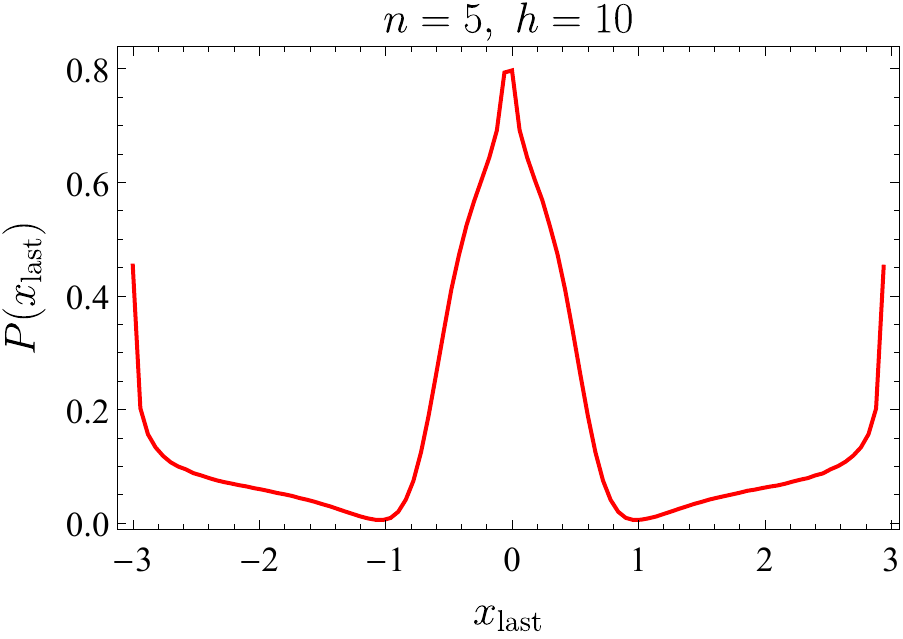}
\caption{Last hitting point densities for the potential~(\ref{eq:potential})  with $n=5,h=1$ (top) and $n=5,h=10$ (bottom).}
\label{fig:lhp}
\end{figure}

The last hitting point, $x_{\mathrm{last}}$, is the last visited point before the escape from the domain $|x|\leqslant 3$
\begin{equation}
    x_{\mathrm{last}}=\{x(t-\Delta t) : |x(t-\Delta t)| \leqslant 3 \land |x(t)|>3\}.
\end{equation}
The last hitting point is a random variable.
Therefore, from an ensemble of points $x_{\mathrm{last}}$, it is possible to estimate the last hitting point density $P(x_{\mathrm{last}})$.
Sample last hitting point densities are depicted in Fig.~\ref{fig:lhp}.

From examination of the last hitting point densities it is distinct that there are two main escape scenarios from the system, see Fig.~\ref{fig:lhp}.
The first scenario is typical for L\'evy flights: a particle might escape from any finite domain by a single very long jump.
This scenario produces a part of the last hitting point density which bears similarities to the quasi-stationary states or time dependent probability densities at long time, see bottom panels of Figs.~\ref{fig:pdfsh1} -- \ref{fig:pdfsh10}.
More precisely, the last hitting point density consists of the single peak superimposed on the quasi-stationary density.
In Fig.~\ref{fig:lhp}, the dominating peak at the origin is associated with the initial condition and its persistence at short times, while the remaining $|x|<2$ part corresponds to the quasi-stationary density.
In the second scenario, a particle jumps out of the potential well but does not manage to escape from the domain.
After leaving the potential well, especially for large $n$, a  particle is almost free as the potential profile is quite flat.
If a particle starts to move along the flat part of the potential it might escape from the system via many short jumps.
Analogously like in the case of escape from finite intervals \cite{capala2020peculiarities}, also here, the sequence of short jumps is responsible for the emergence of maxima of the last hitting point densities near the boundaries, because short jumps give the highest chances to approach the absorbing boundary.

Fig.~\ref{fig:lhp} presents histograms of the last hitting point for two exemplary sets of parameters: $n$ (barrier width) and $h$ (depth of the potential well).
In the top panel, which depicts results for $n=5$ and $h=1$, one may observe the central dominating peak and two smaller ones.
The central peak corresponds to the initial condition while the smaller peaks correspond to the modes of the quasi-stationary density which have emerged in the potential well.
Moreover, in the vicinity of the absorbing boundaries there are two modes in the last hitting point density produced by escapes via series of jumps.
The bottom panel of Fig.~\ref{fig:lhp} displays results for $n=5$ and $h=10$, for which the inner ($|x|<1$) part of the time dependent density is always unimodal, see Fig.~\ref{fig:pdfsh10}.
Therefore, the last hitting point density is composed of a narrow peak in the central part corresponding to particles which have escaped via a single jump from the potential well and two outer peaks in the proximity of the absorbing boundaries.

\subsection{First passage times\label{sec:fpt}}

\begin{figure}[H]
\includegraphics[width=0.9\columnwidth]{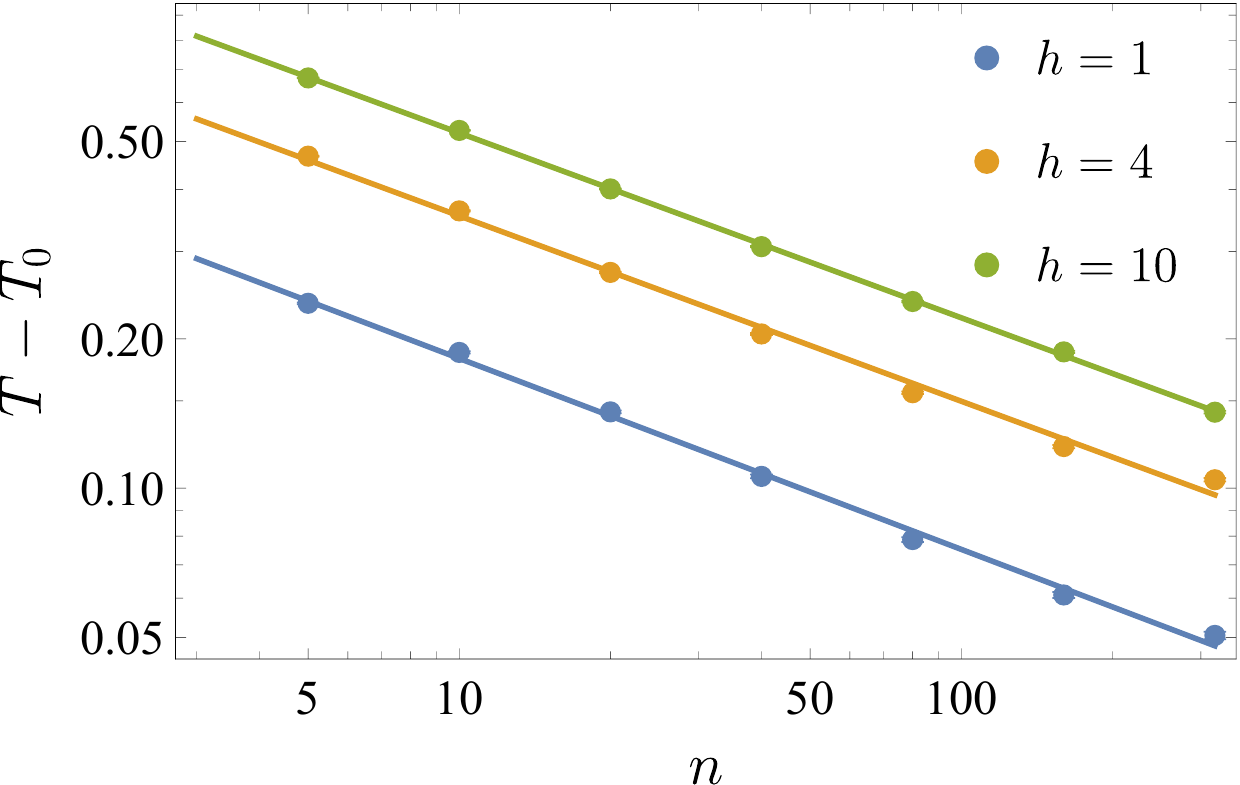}
\caption{Difference between mean first passage times $T$ for the considered potential and \bdt{ $T_0$ for the free particle,} given by Eq.~(\ref{eq:MFPTfree}) with $L=3$, as a function of the barrier width ($n$) for different depths  ($h$) of potential wells.
Solid lines correspond to the power-law fits given by Eq.~(\ref{eq:fit}).}
\label{fig:mfpt}
\end{figure}

The first passage time $\tau$ is defined as
\begin{equation}
    \tau=\min(t : x(0)=0 \land |x(t)| \geqslant 3)
\end{equation}
while the mean first passage time (MFPT) is the \bdt{ensemble} average first passage time, i.e., $T=\langle \tau \rangle$.

From Fig.~\ref{fig:mfpt} it is clearly visible that with increasing steepness of the potential barrier (increasing $n$) the MFPT  is reduced.
For finite barrier width, $n$, the MFPT is also sensitive to the barrier height $h$.
\textcolor{black}{From computer simulations, we see that} in the limit of $n\to\infty$ the MFPT becomes insensitive \textcolor{black}{or weakly sensitive} to the barrier height, see Refs.~\onlinecite{ditlevsen1999,bier2018,capala2020peculiarities}.
\textcolor{black}{The lack of sensitivity to the barrier height can be intuitively justified by the fact that with the increasing $n$ width of the barrier decreases making the domain where the deterministic force acts narrower.
Consequently, only a very limited number of particles feel the deterministic force, while the majority of particles moves practically as free particles.}
Moreover, the MFPT is dominated by the flat parts of the potential.
More precisely, the escape process is constituted of three phases: approaching the barrier ($|x|=1$), passing over the barrier, and approaching the absorbing boundaries ($x=\pm 3$).
For $n\to\infty$, with any finite $h$, the width of the potential barrier tends to zero.
The transition over the ``zero'' width potential barrier is immediate, thus the process of approaching the barriers and boundaries determines the value of the MFPT.
Finally, for any finite $h$, in the limit of $n\to\infty$, the MFPTs are the same.
Moreover, for $n\to\infty$ the MFPT tends to the MFPT \cite{blumenthal1961,getoor1961} of a free particle from a finite domain  of half-width $L$,
\begin{equation}
    T_0=\frac{1}{\Gamma(1+\alpha)} \frac{L^\alpha}{\sigma^\alpha}.
    \label{eq:MFPTfree}
\end{equation}
For the given setup \bdt{$L=3$ and} the MFPT is equal to $T_0=3$.
Additionally, in Fig.~\ref{fig:mfpt}, to the numerically obtained values of the differences between MFPTs and $T_0$, power-law functions
\begin{equation}
T - T_0 = b \times x^{-a}
    \label{eq:fit}
\end{equation}
have been fitted and depicted by solid lines.
Values of the obtained parameters are included in Tab.~\ref{tab:exponents}.
Note that the exponents $a$ corresponding to different potential barrier heights are similar.

\begin{table}[!h]
\begin{tabular}{c||c|c|c}
$h$ & 1 & 4 & 10  \\ \hline\hline
$a$ & 0.385 & 0.374 & 0.373 \\
$b$ & 0.44 & 0.84 & 1.23 \\
\end{tabular}
\caption{Values of the fitted parameter in Eq.~(\ref{eq:fit}).}
\label{tab:exponents}
\end{table}

\begin{figure}[H]
    \centering
    \includegraphics[width=0.9\columnwidth]{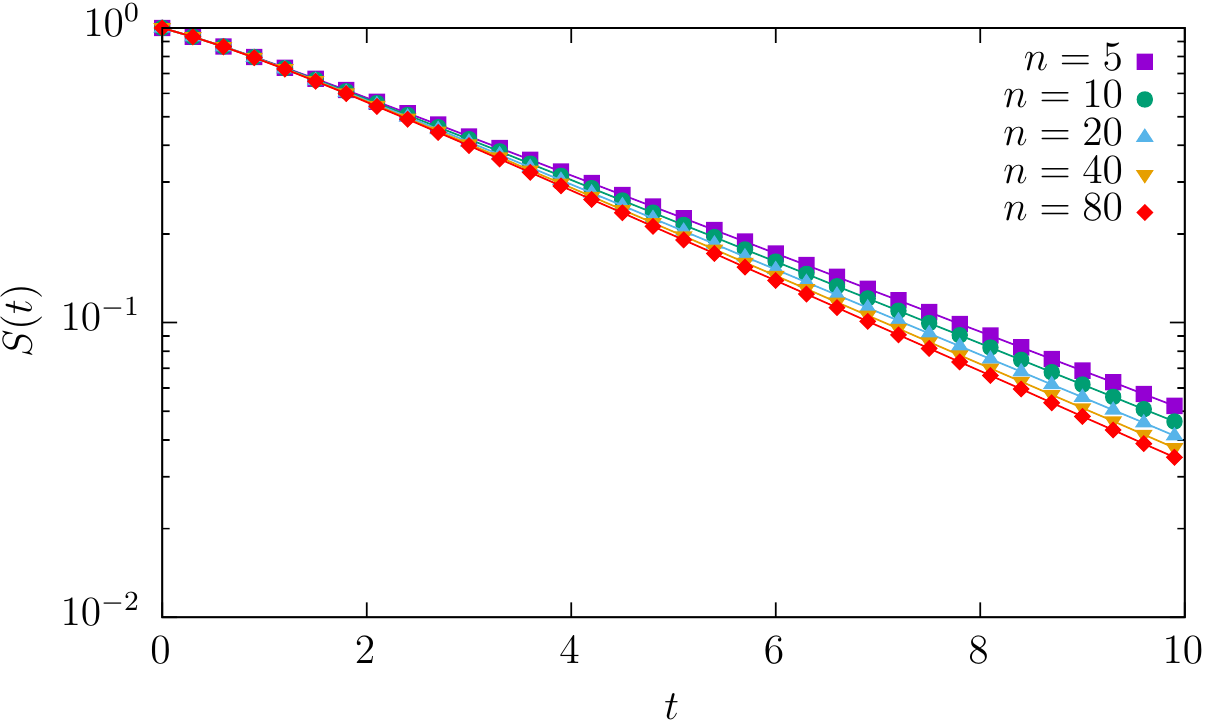}\\
    \includegraphics[width=0.9\columnwidth]{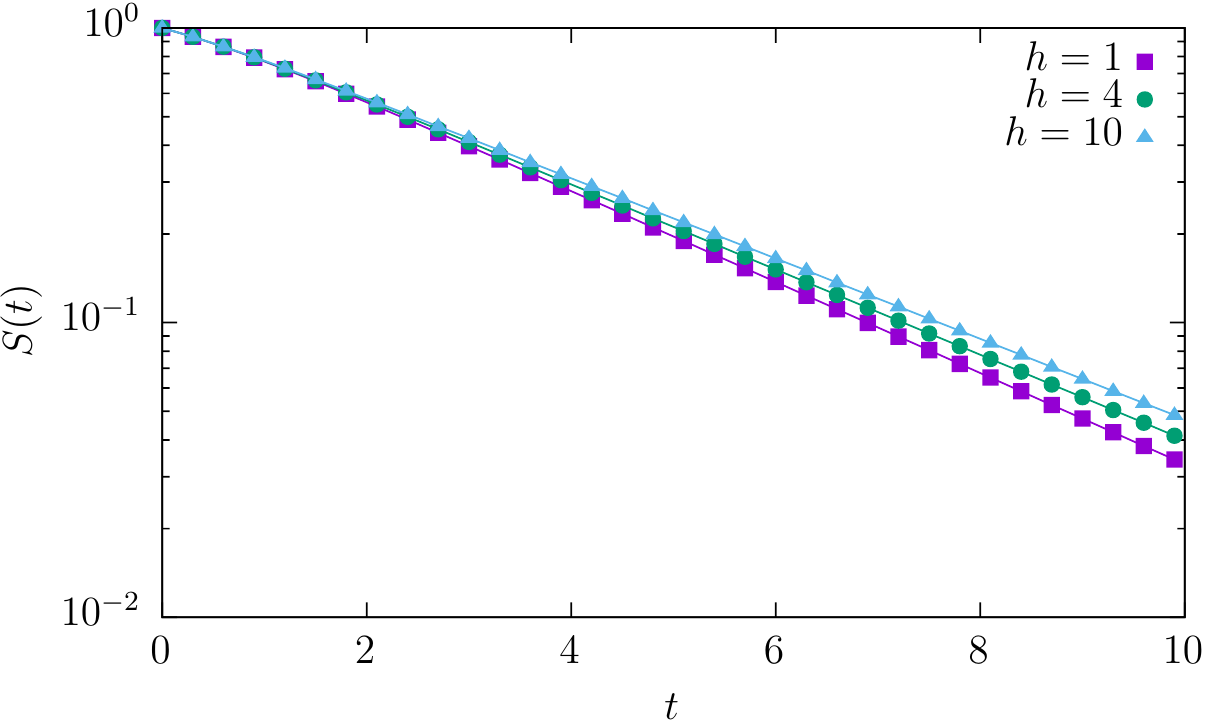}
    \caption{The survival probability $S(t)$, i.e., the probability that a particle has not escaped from the potential well up to time $t$ for fixed $h=4$ with various $n$ (top panel) and for fixed $n=20$ with varying $h$ (bottom panel).}
    \label{fig:fpt}
\end{figure}

The process $x(t)$, see Eq.~(\ref{eq:langevin}), is Markovian.
Consequently, the survival probability $S(t)$ given by Eq.~(\ref{eq:survival}) has exponential tails $S(t)\propto \exp(-\lambda t)$, see Fig.~\ref{fig:fpt} and \cite{dybiec2017levy}.
The survival probability $S(t)$ is related to the first passage time density $\wp(t)$ via $\wp(t)=-\frac{dS(t)}{dt}$.
At short times, the survival probabilities $S(t)$ corresponding to various parameters are similar, see~Fig.~\ref{fig:fpt}, because those particles which have not escaped via a single jump need some time to approach the boundary.
The differences between various $S(t)$ show up at longer times $t$, because particles which have stayed longer in the system were capable of penetrating the potential barrier.

Top panel of Fig.~\ref{fig:fpt} shows survival probabilities for the fixed height $h=4$ of the potential barrier and varying width of the barrier $n$ ($n\in\{5,10,20,40,80\}$).
The bottom panel of Fig.~\ref{fig:fpt} depicts survival probabilities for the fixed width $n=20$ of the potential barrier and varying height of the barrier $h$ ($h\in\{1,4,10\}$).
The decay rate of survival probabilities is related to the mean first passage time \cite{dybiec2017levy}.
For fixed $n$ with the increasing $h$ the MFPT increases, while for fixed $h$ with increasing $n$ MFPT decreases.
With decreasing MFPT the decay of the survival probability becomes faster, i.e., the exponent $\lambda$ characterizing the exponential decay increases \cite{dybiec2017levy}, see Fig.~\ref{fig:fpt}.

%%%%%%%%%%%%%%%%%%%%%%%%%%%%%%%%%%%%%%%%%%%%%%%%%%%%%%%%%%
\section{Time dependent densities for the case $\alpha\neq1$ \label{sec:neq1}}

\begin{figure}[h!]
\centering
\includegraphics[width=0.49\columnwidth]{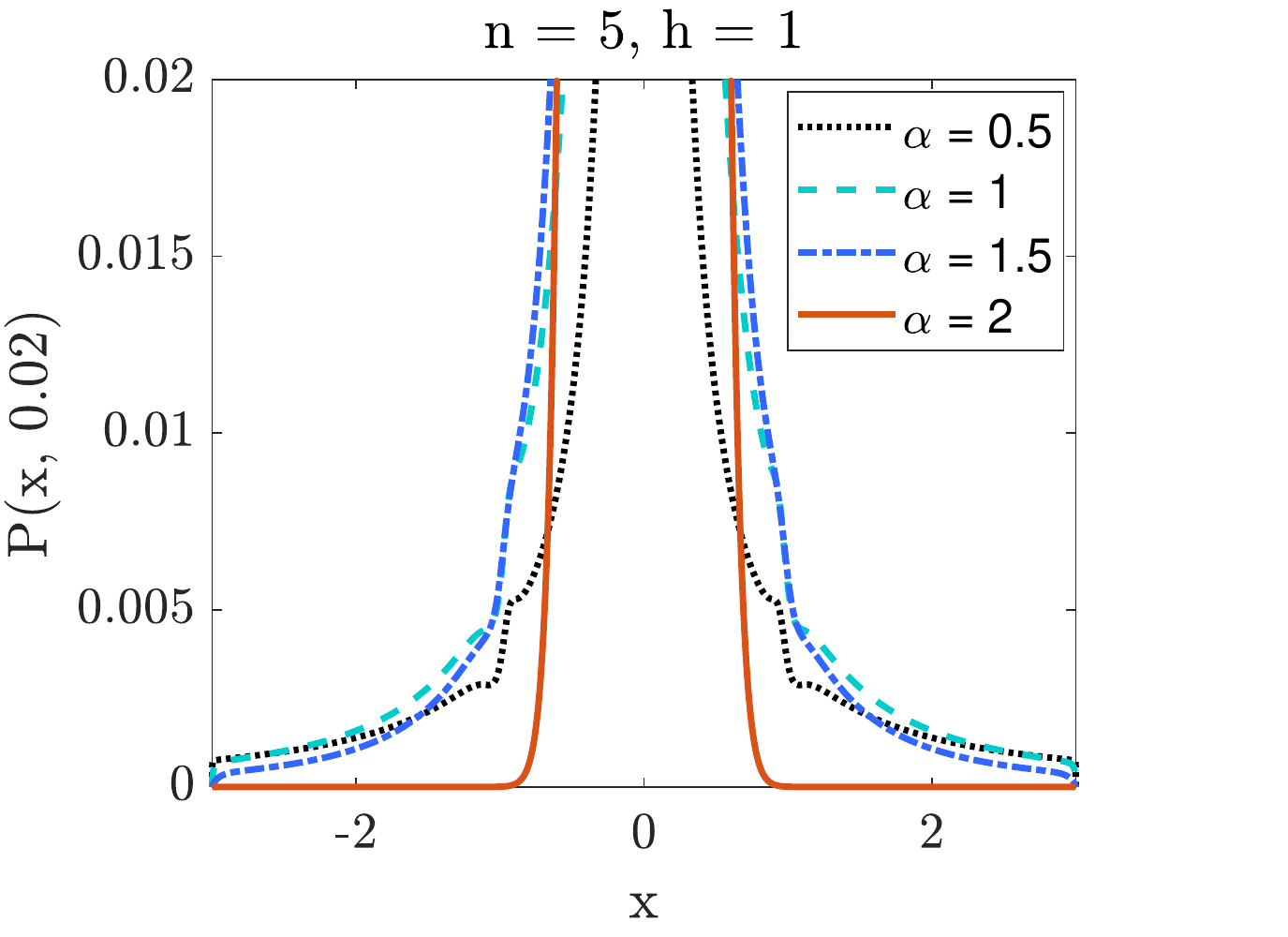}
\includegraphics[width=0.49\columnwidth]{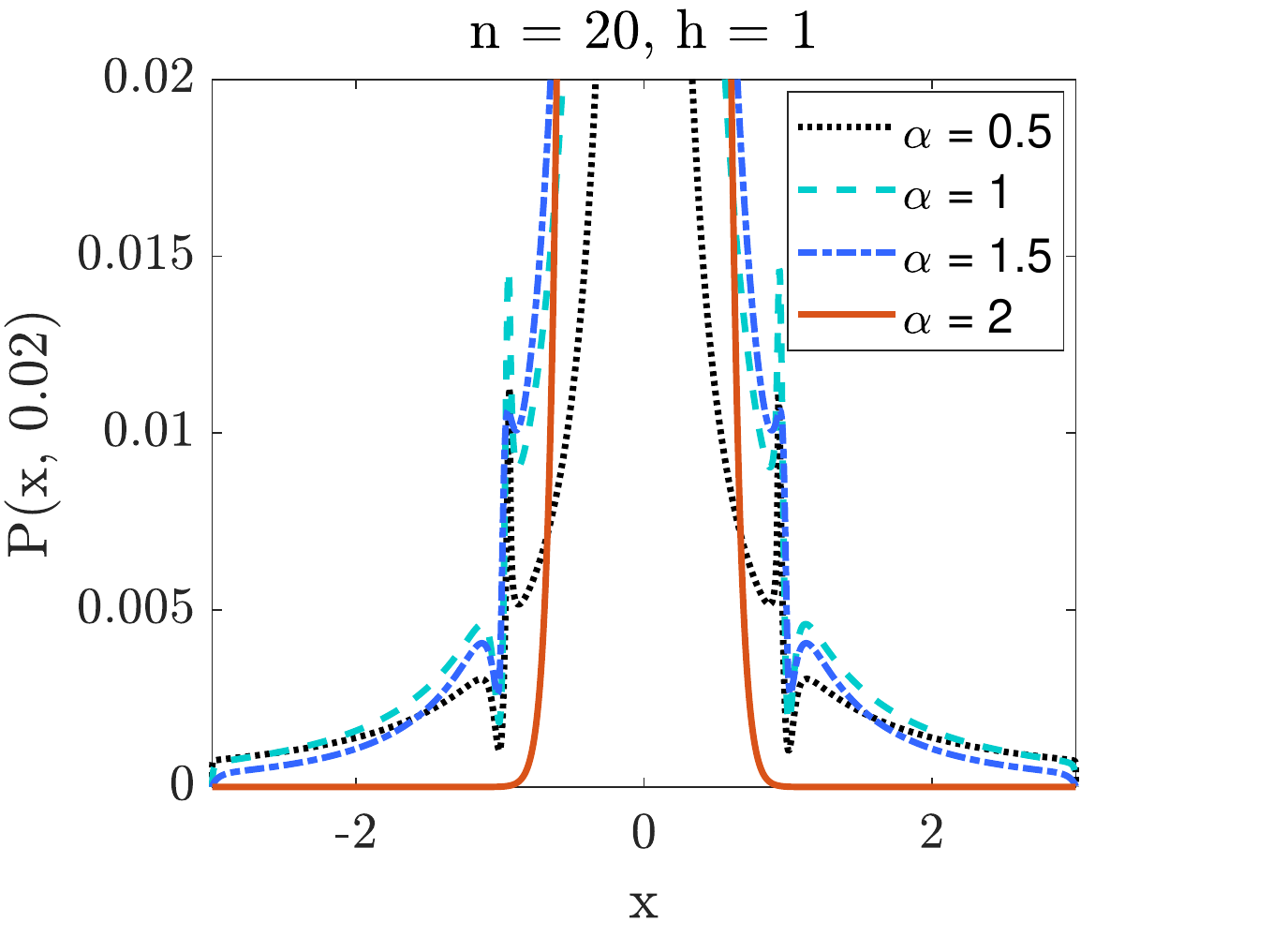}\\
\includegraphics[width=0.49\columnwidth]{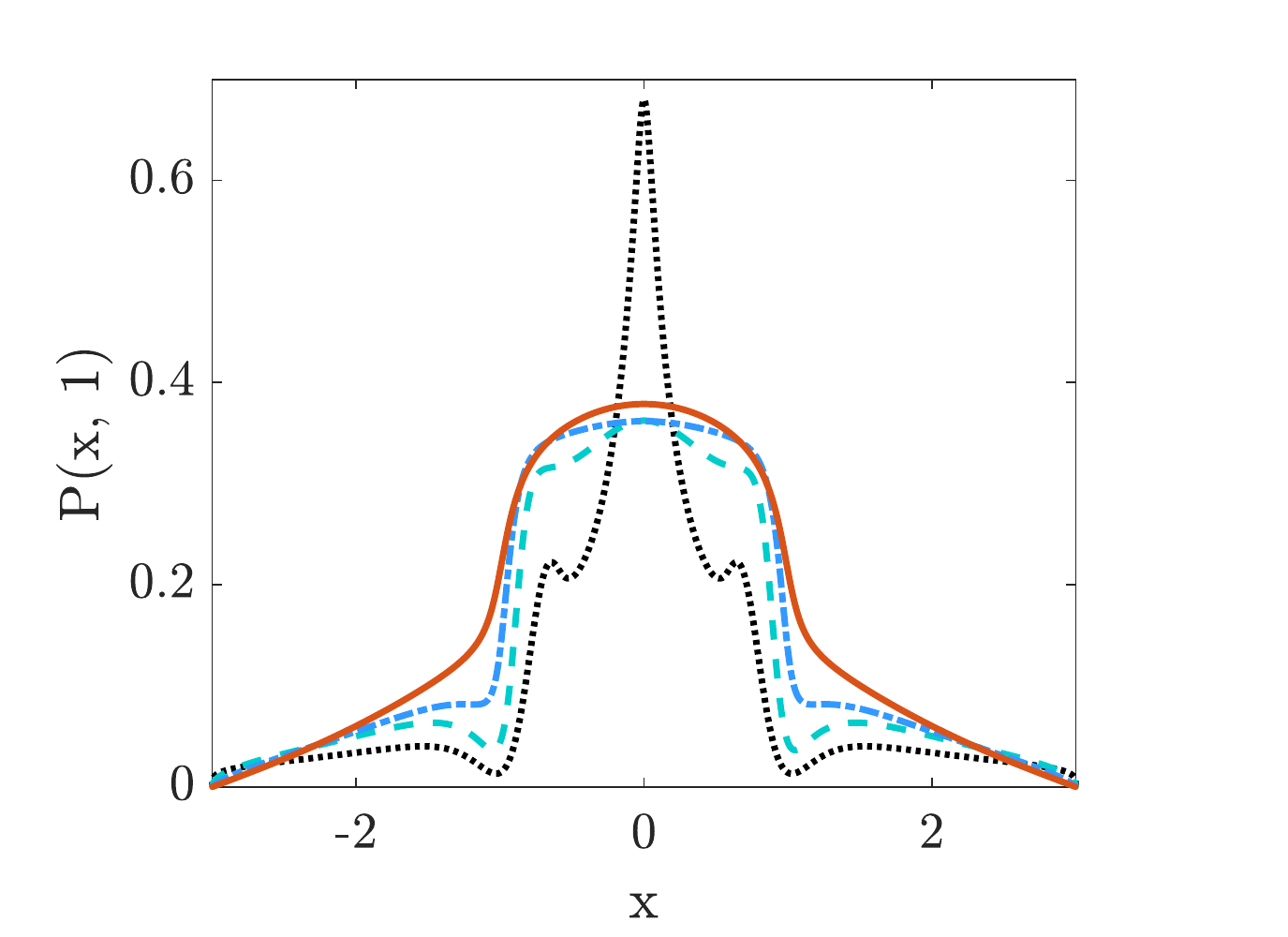}
\includegraphics[width=0.49\columnwidth]{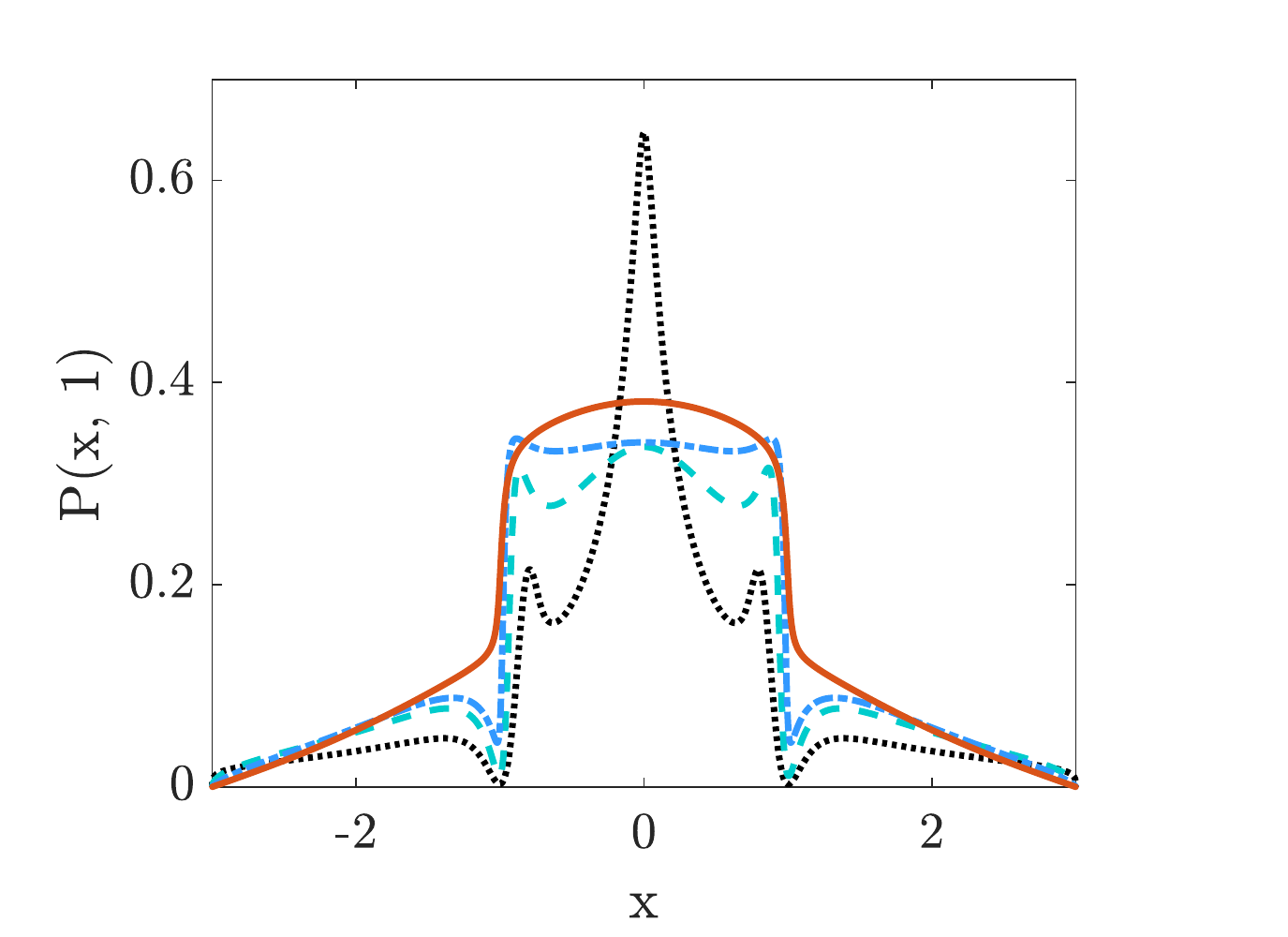}\\
\includegraphics[width=0.49\columnwidth]{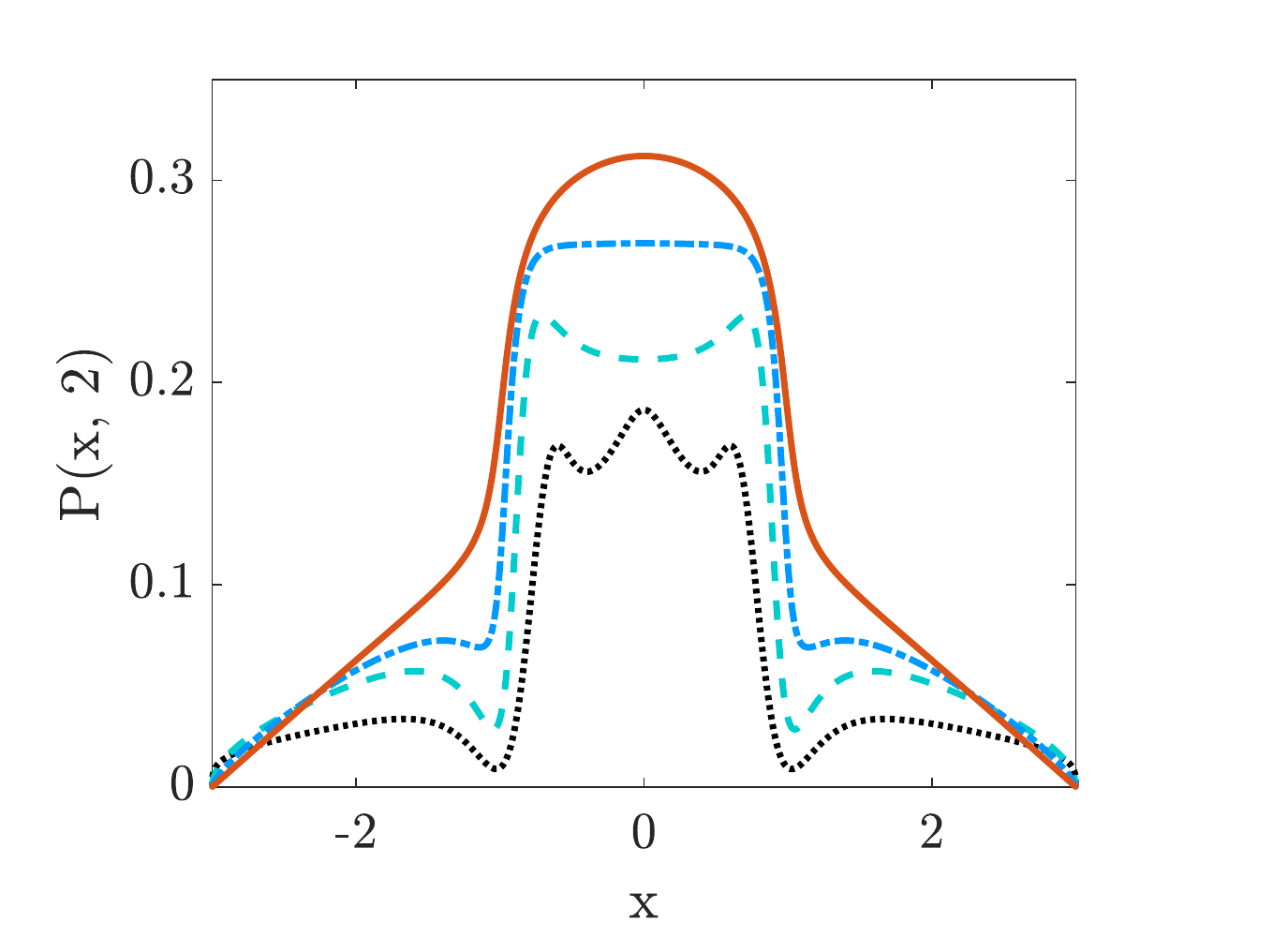}
\includegraphics[width=0.49\columnwidth]{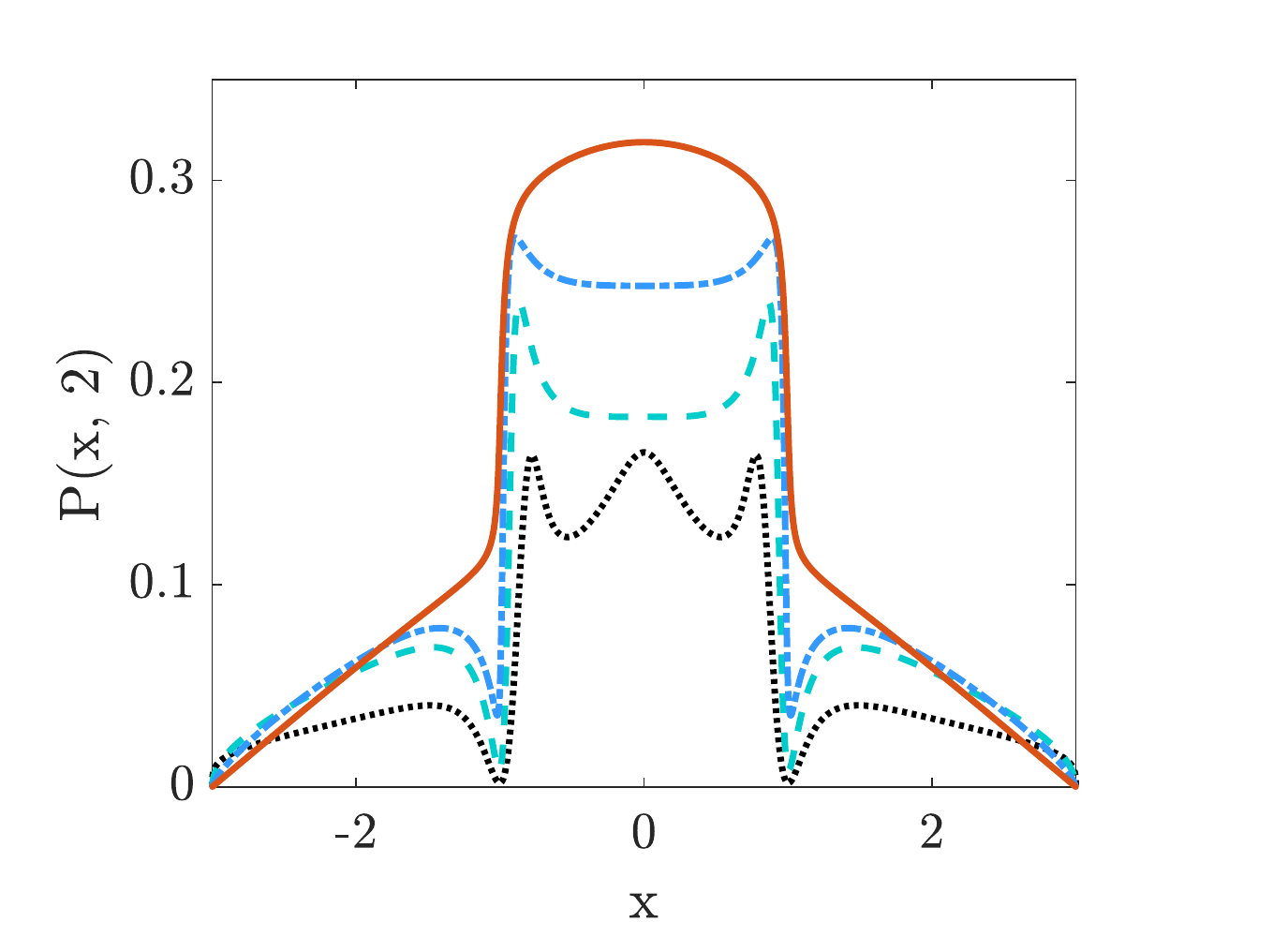}\\
\includegraphics[width=0.49\columnwidth]{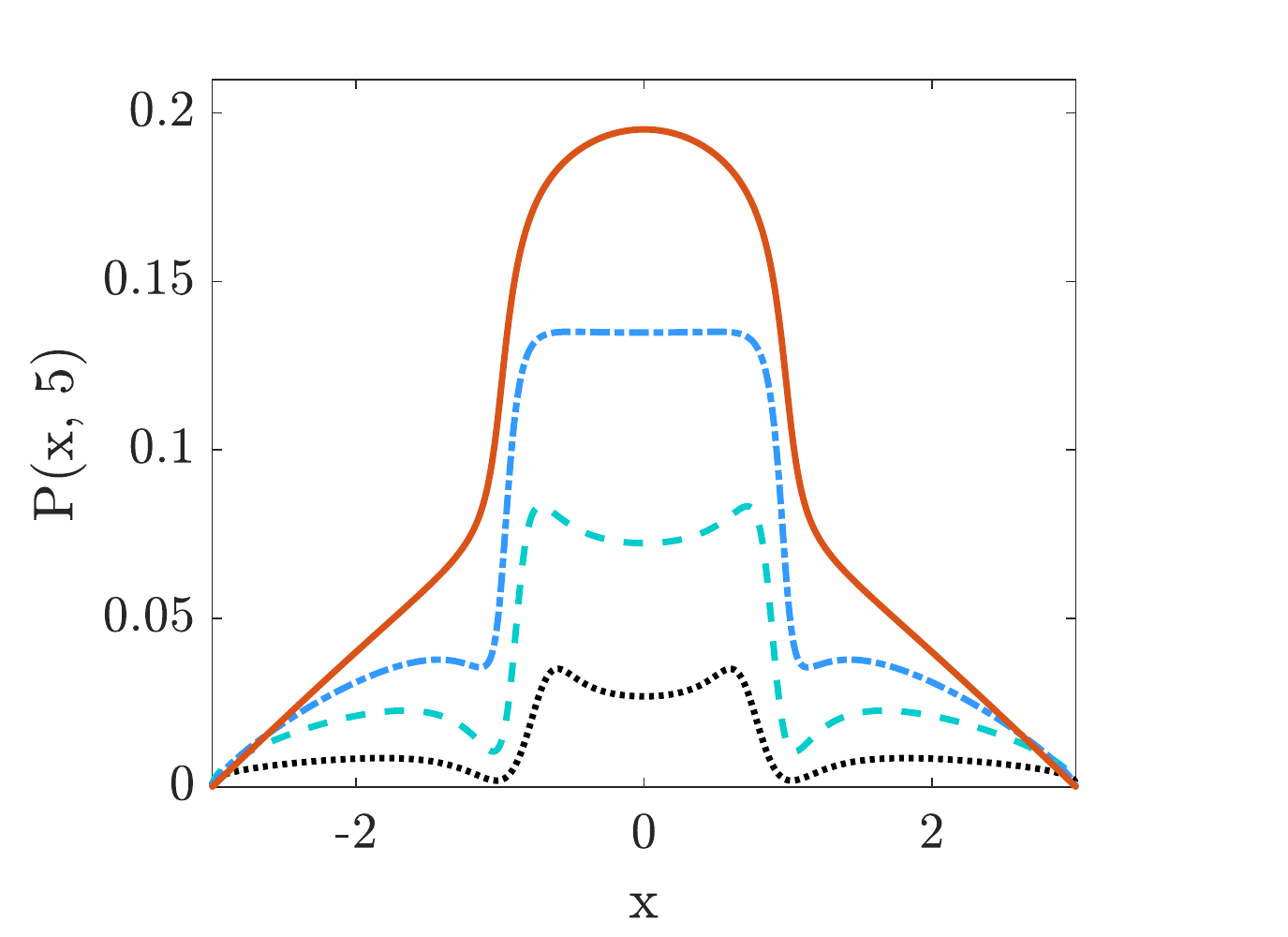}
\includegraphics[width=0.49\columnwidth]{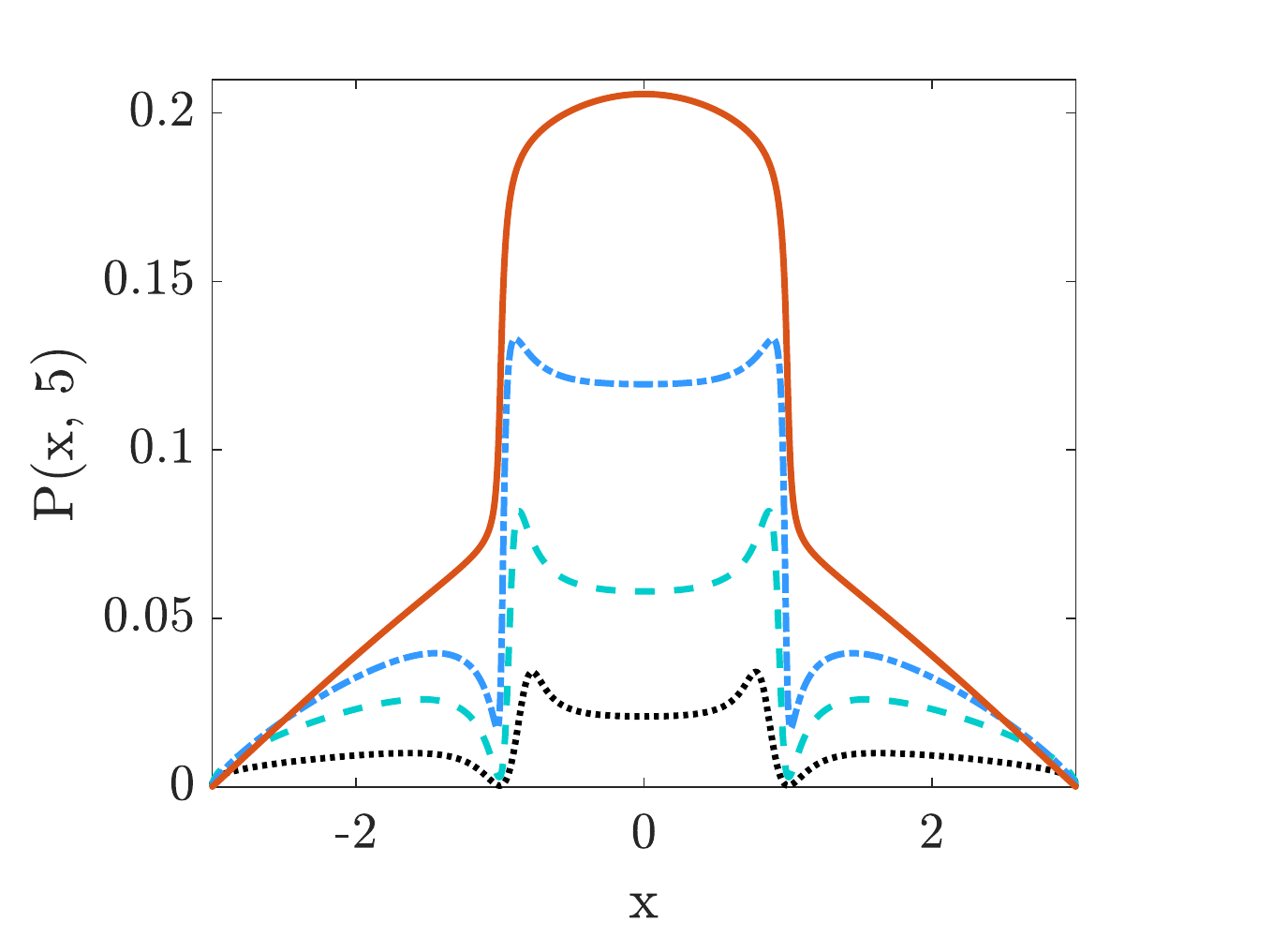}\\
\includegraphics[width=0.49\columnwidth]{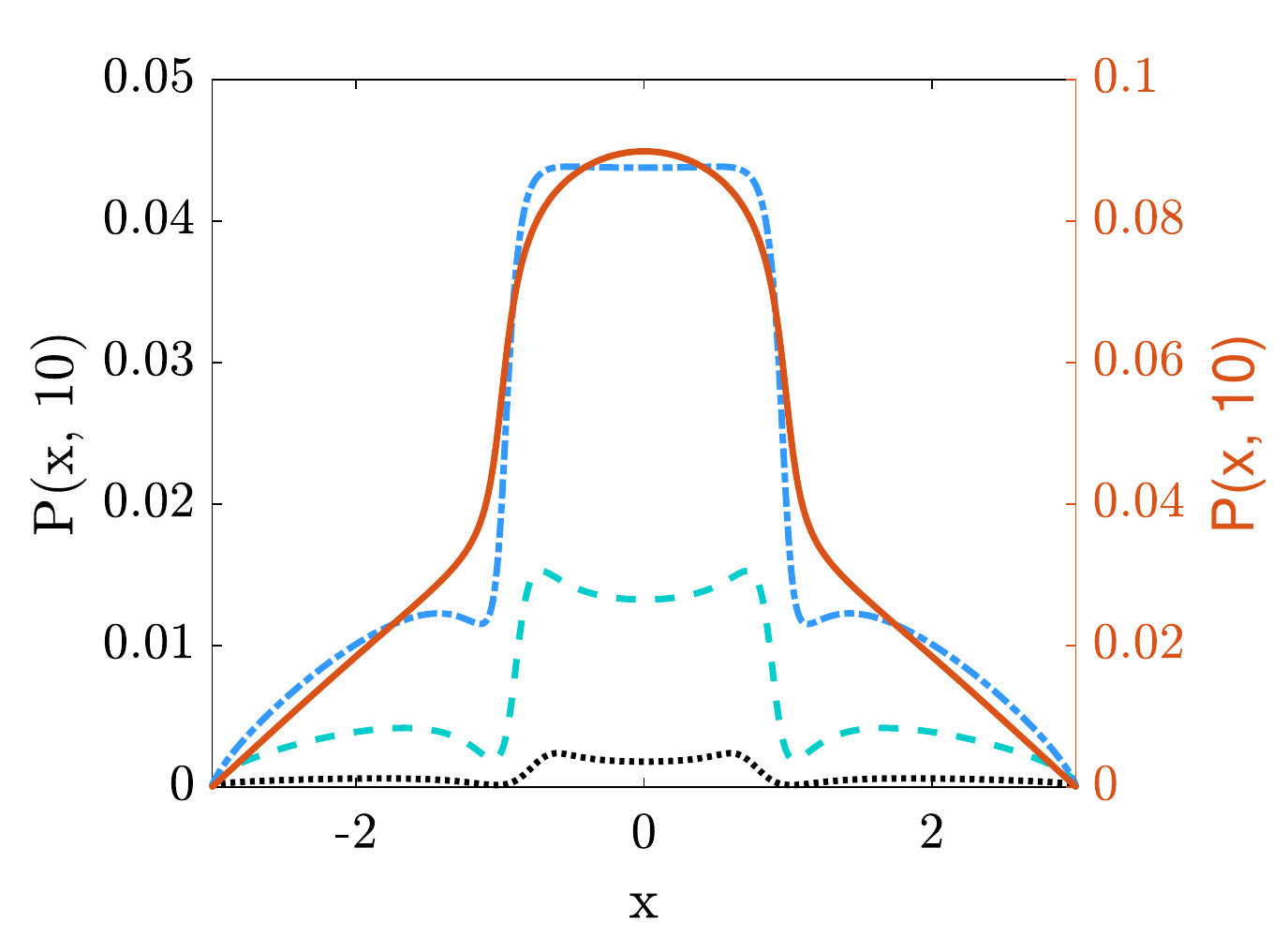}
\includegraphics[width=0.49\columnwidth]{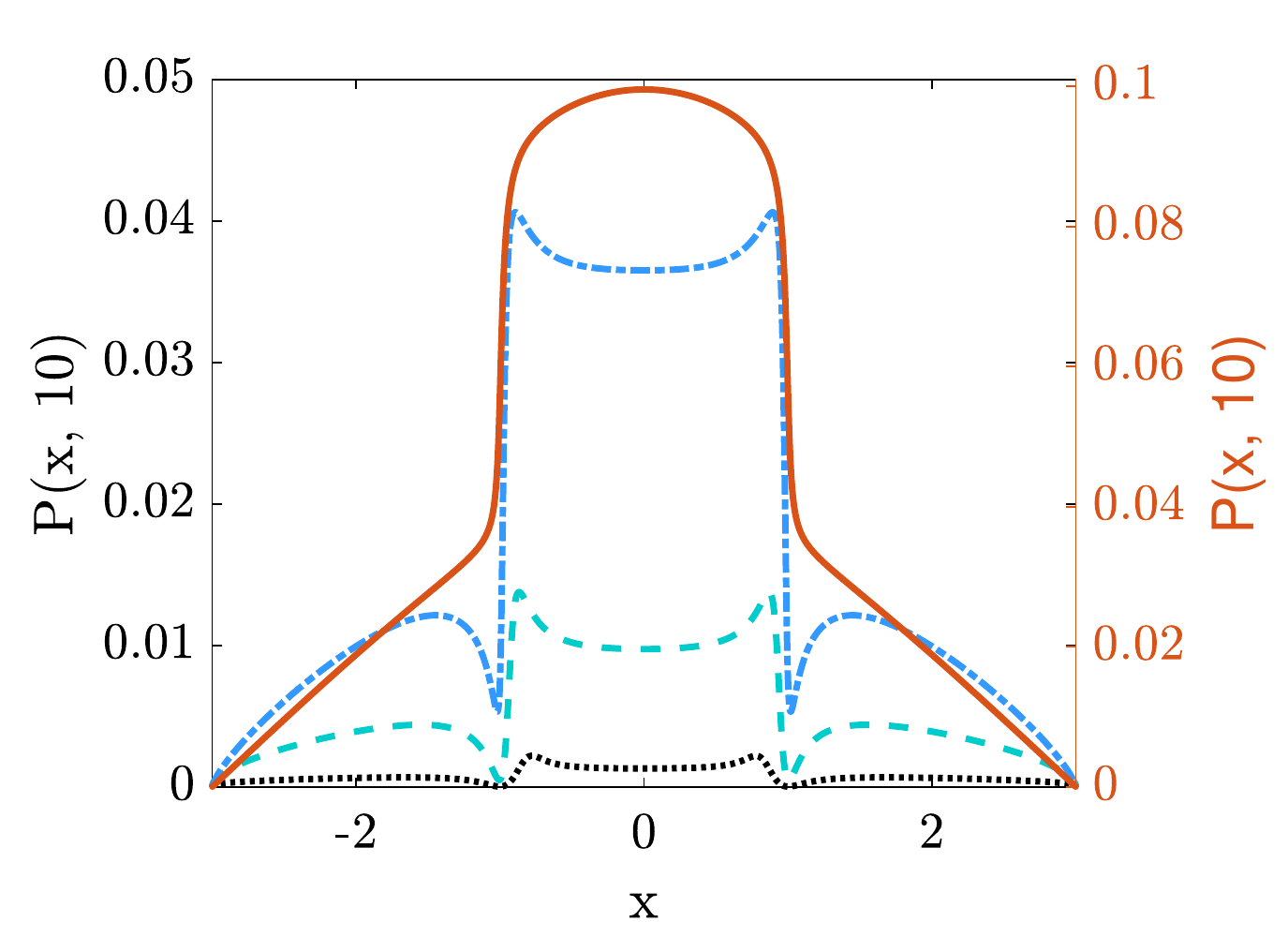}\\
\caption{Time dependent PDFs for various parameters $n$ and $h$ characterizing the \bdt{arctangent} potential (\ref{eq:potential}) at different times $t=0.02$, 1, 2, 5, and $10$ (top to bottom). The columns correspond to $n=5$ (left) and $n=20$ (right).
Results are obtained by numerical solution of the space-fractional Smoluchowski-Fokker-Planck equation (\ref{eq:ffp}) with time step $\Delta t=0.0001$ and space increment $\Delta x=0.001$.
\bdt{For clarity the (unimodal) peaks of time dependent densities in the top panel are not shown, while in the bottom panel the densities for $\alpha=2$ are plotted using the right vertical axis.
}
}
\label{fig:fig1}
\end{figure}

\begin{figure}[h!]
\centering
\includegraphics[width=0.49\columnwidth]{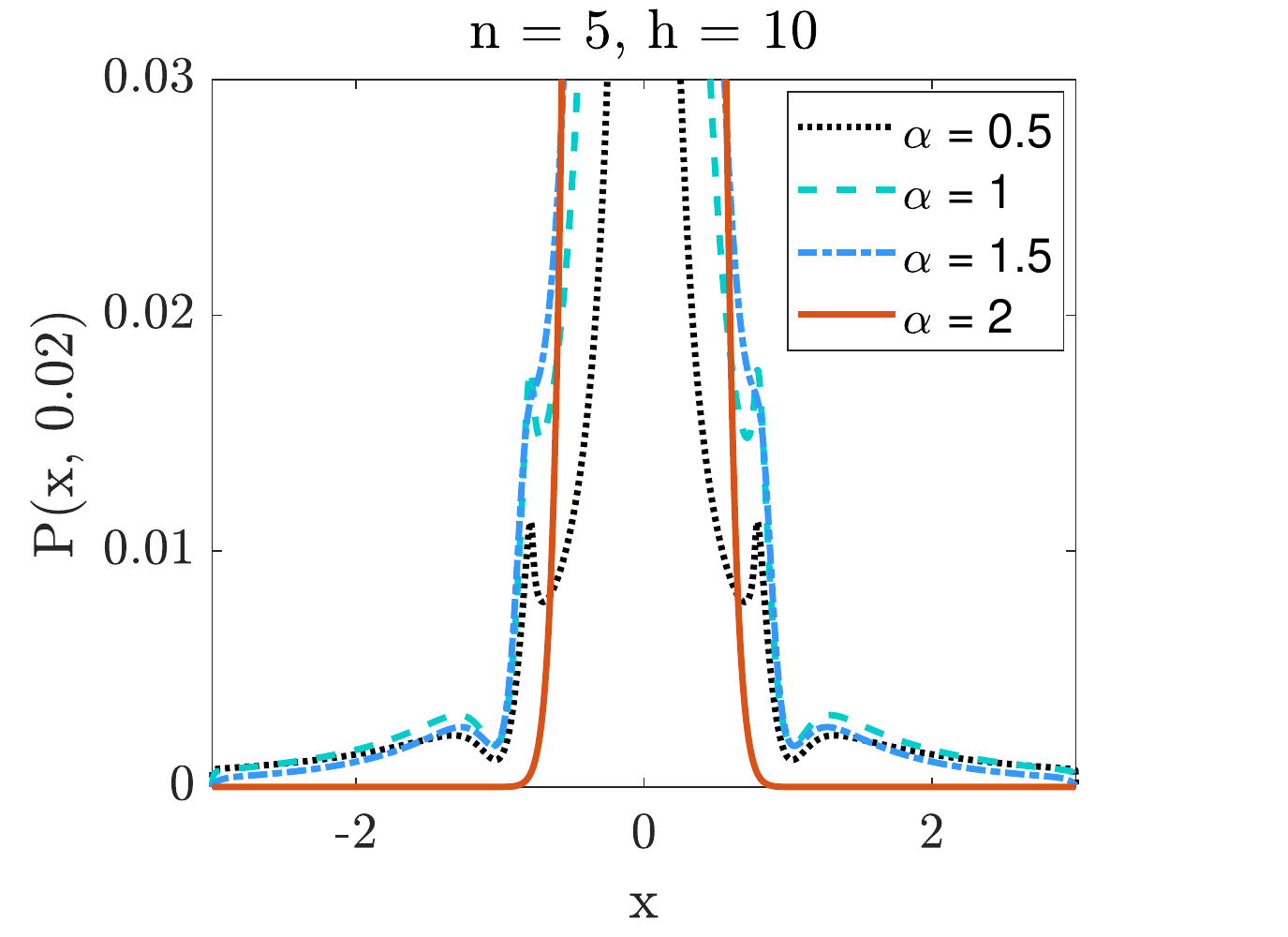}
\includegraphics[width=0.49\columnwidth]{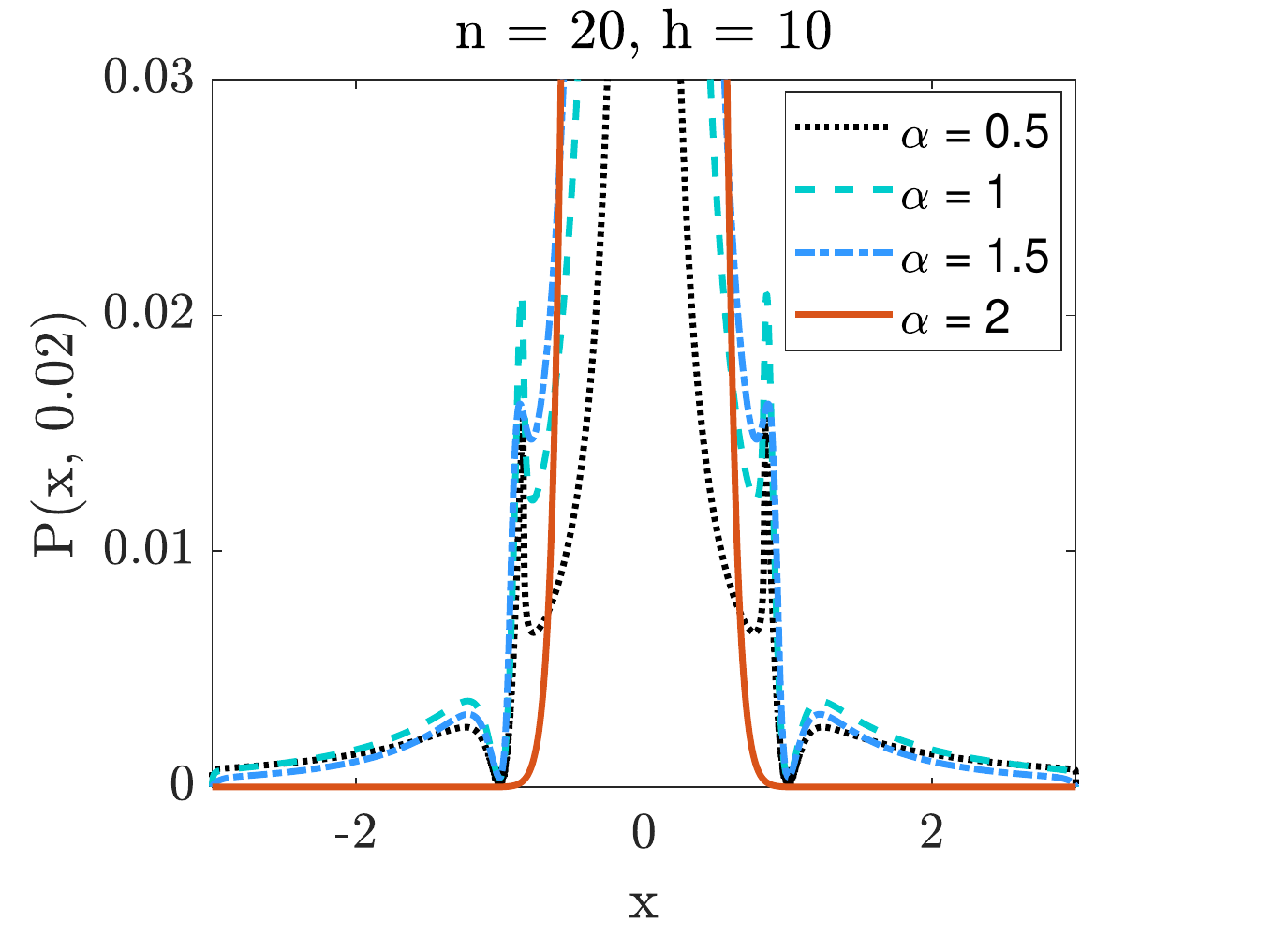}\\
\includegraphics[width=0.49\columnwidth]{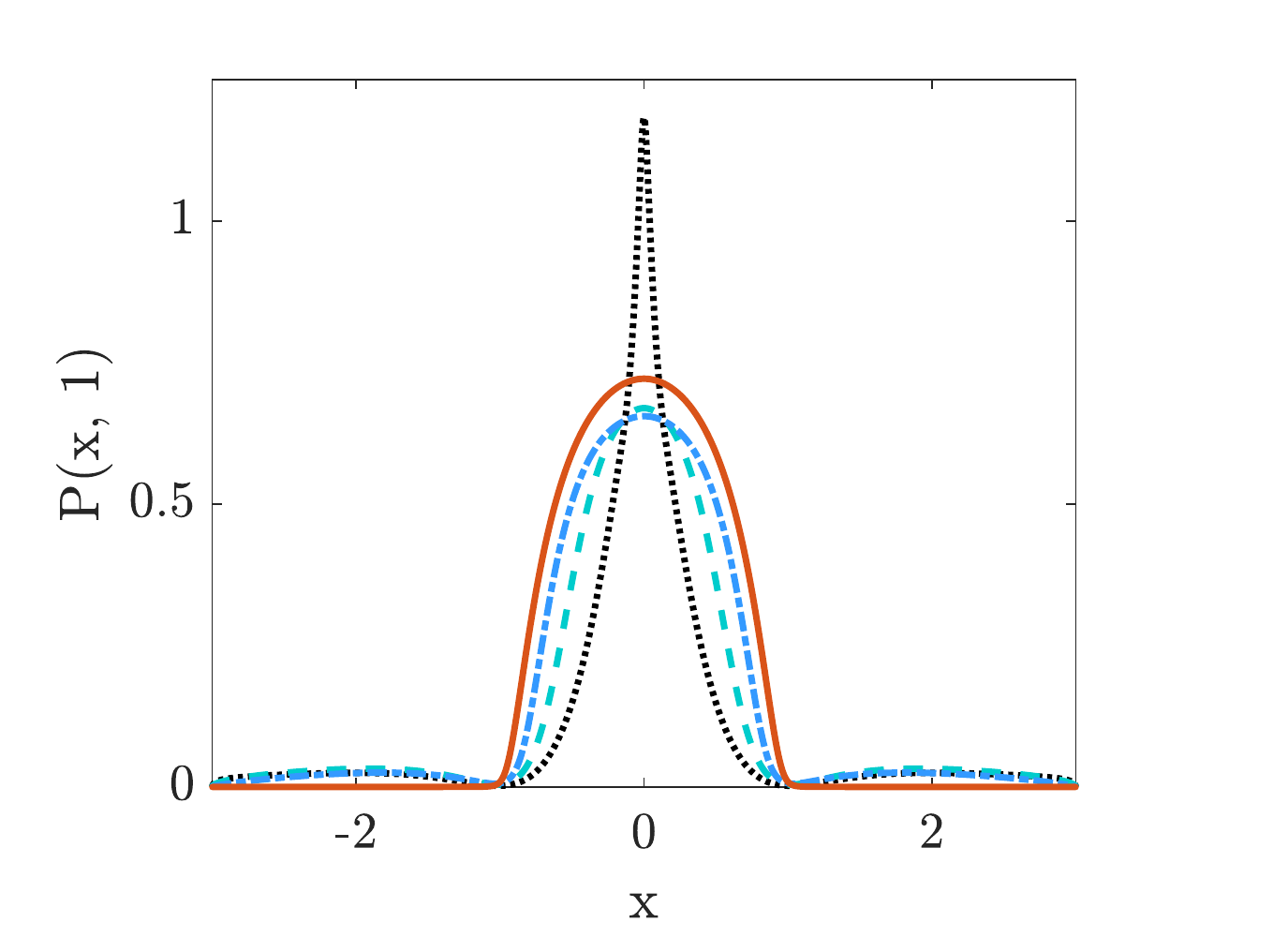}
\includegraphics[width=0.49\columnwidth]{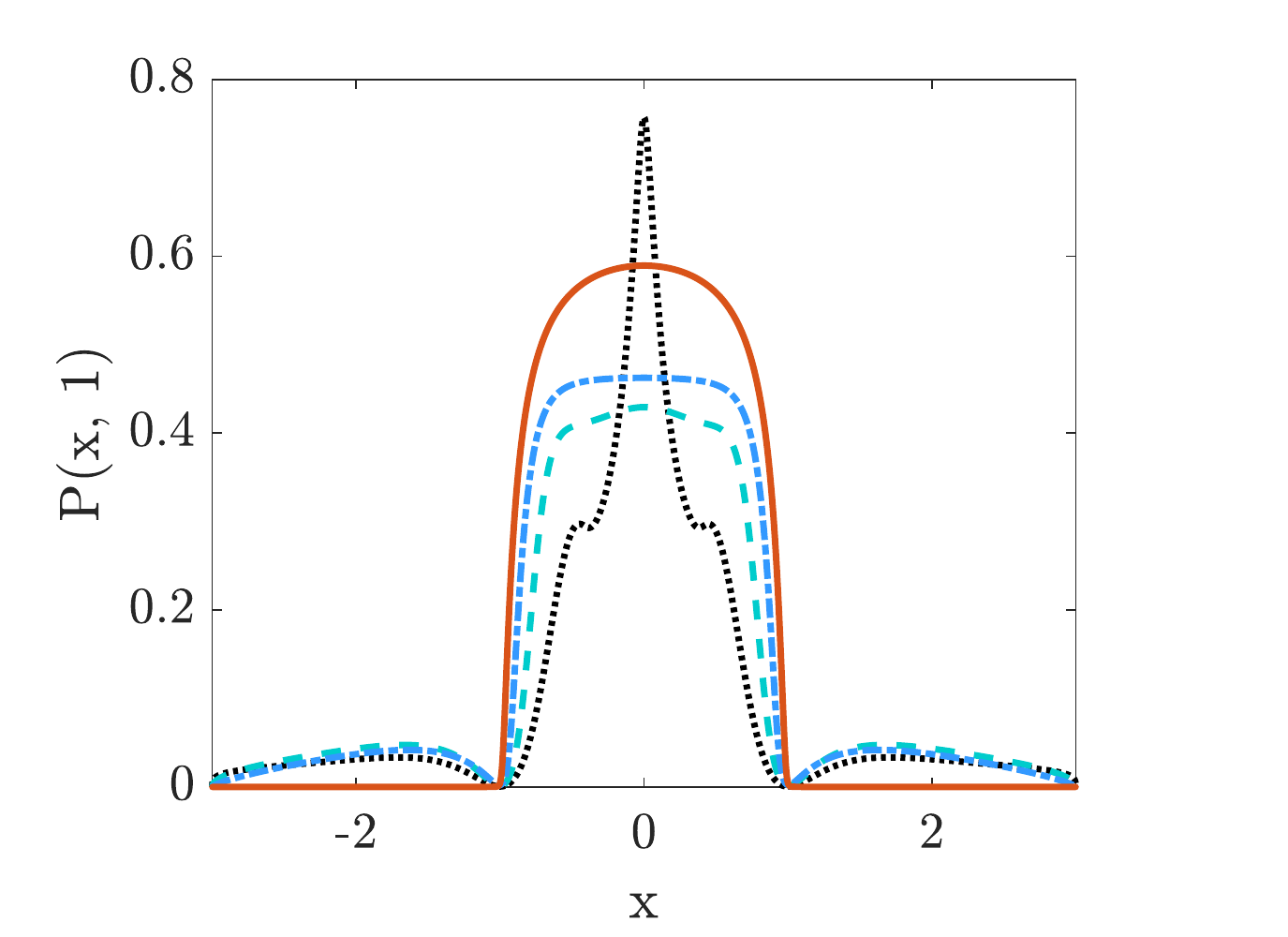}\\
\includegraphics[width=0.49\columnwidth]{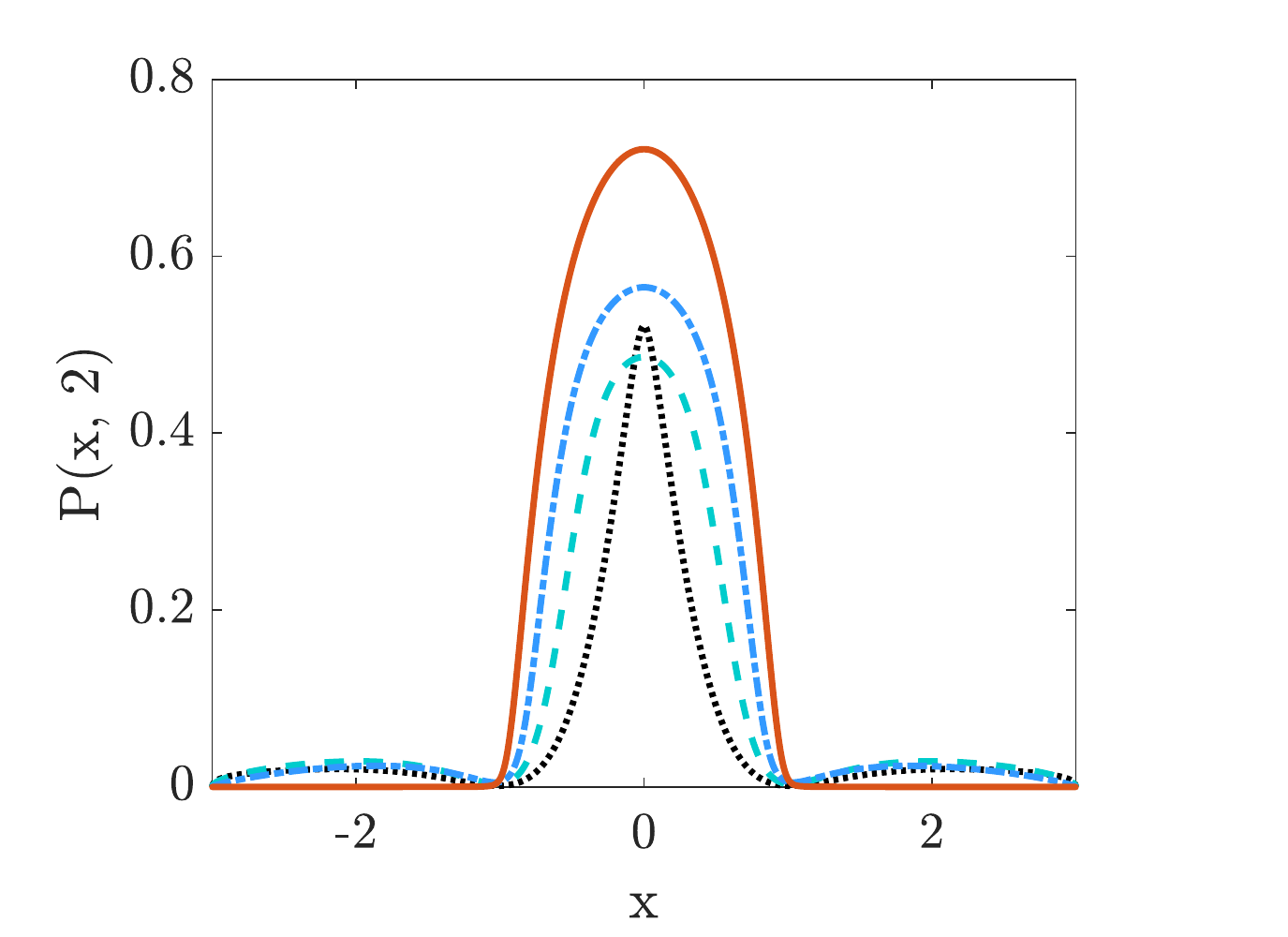}
\includegraphics[width=0.49\columnwidth]{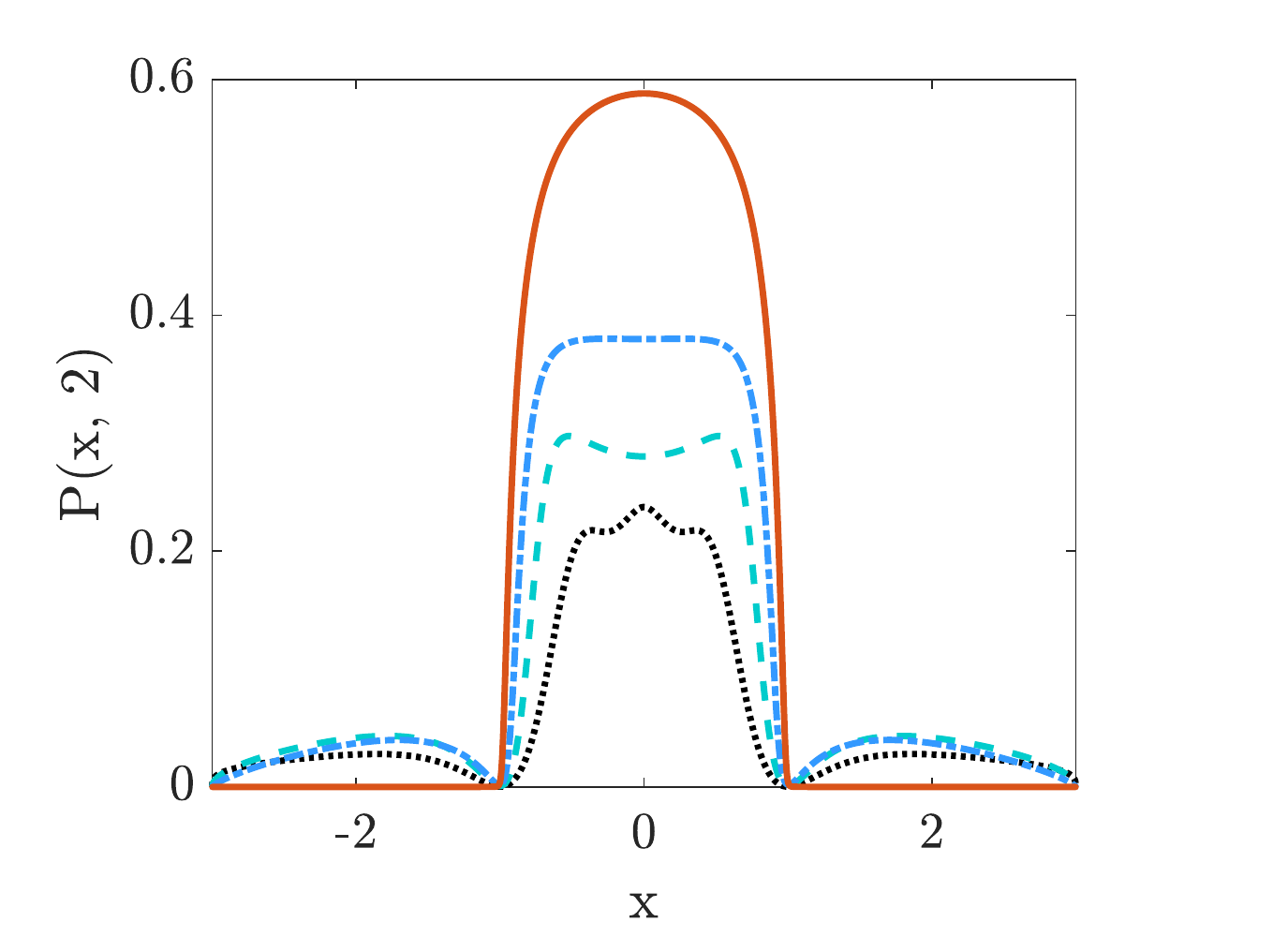}\\
\includegraphics[width=0.49\columnwidth]{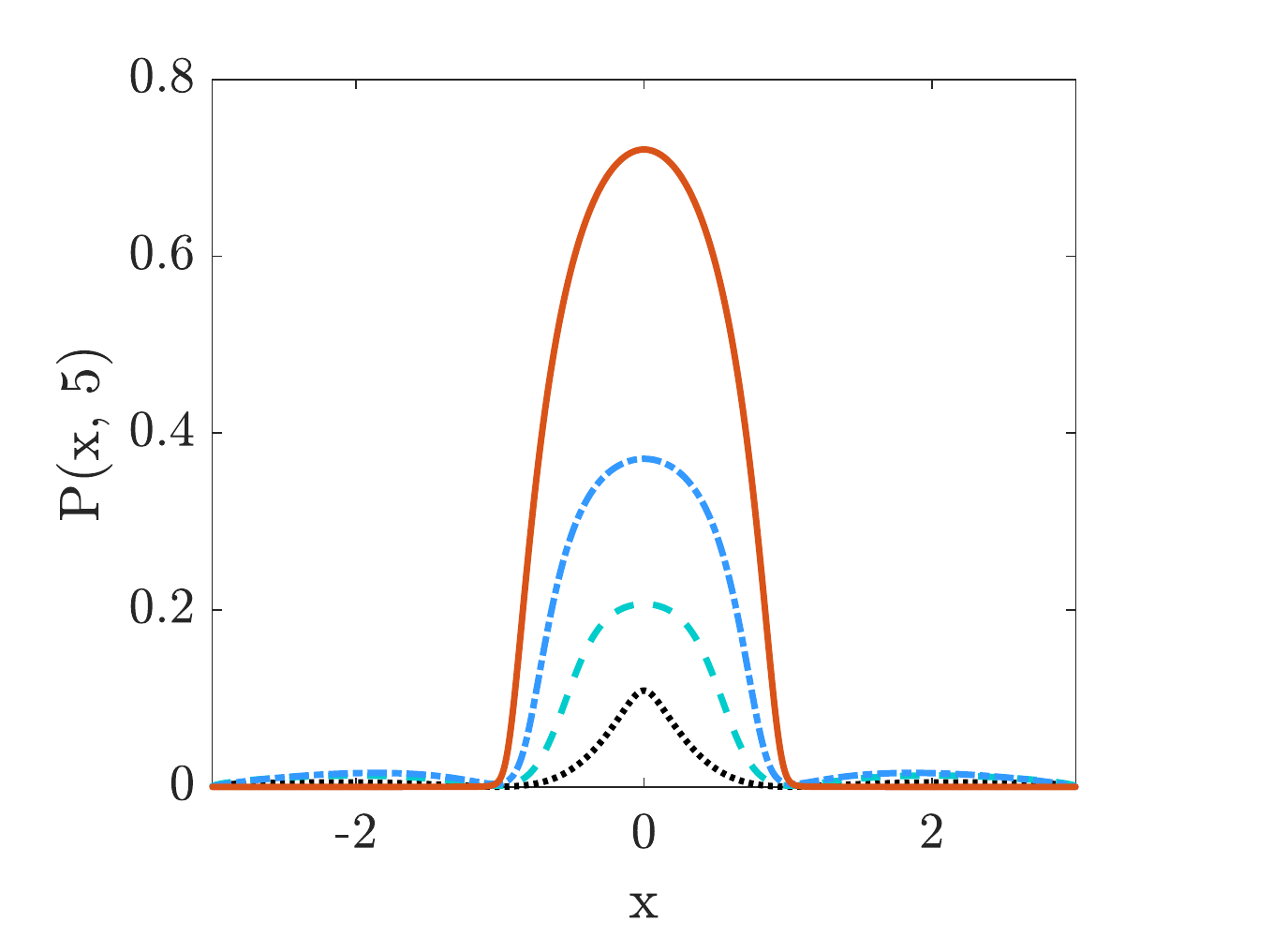}
\includegraphics[width=0.49\columnwidth]{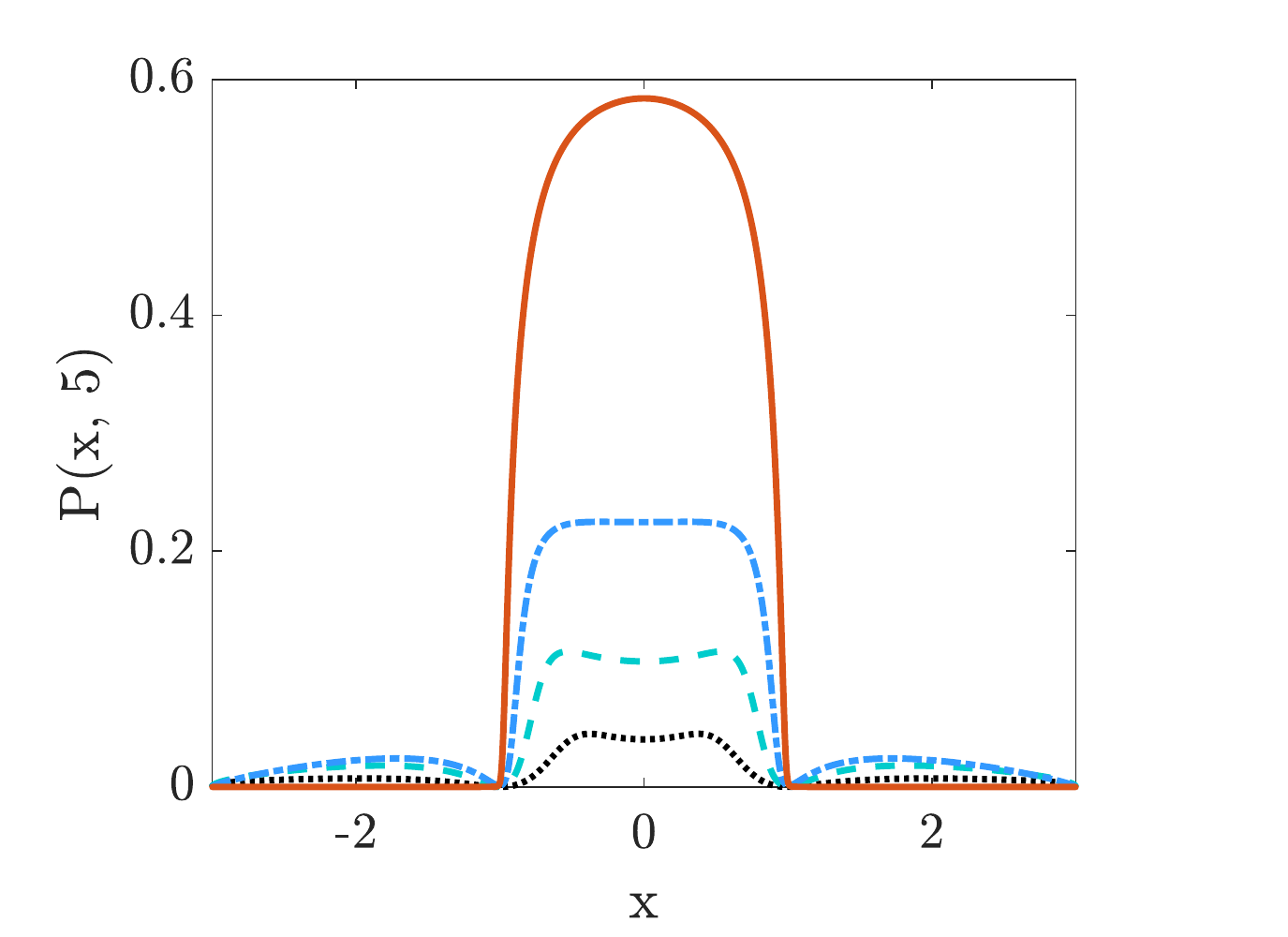}\\
\includegraphics[width=0.49\columnwidth]{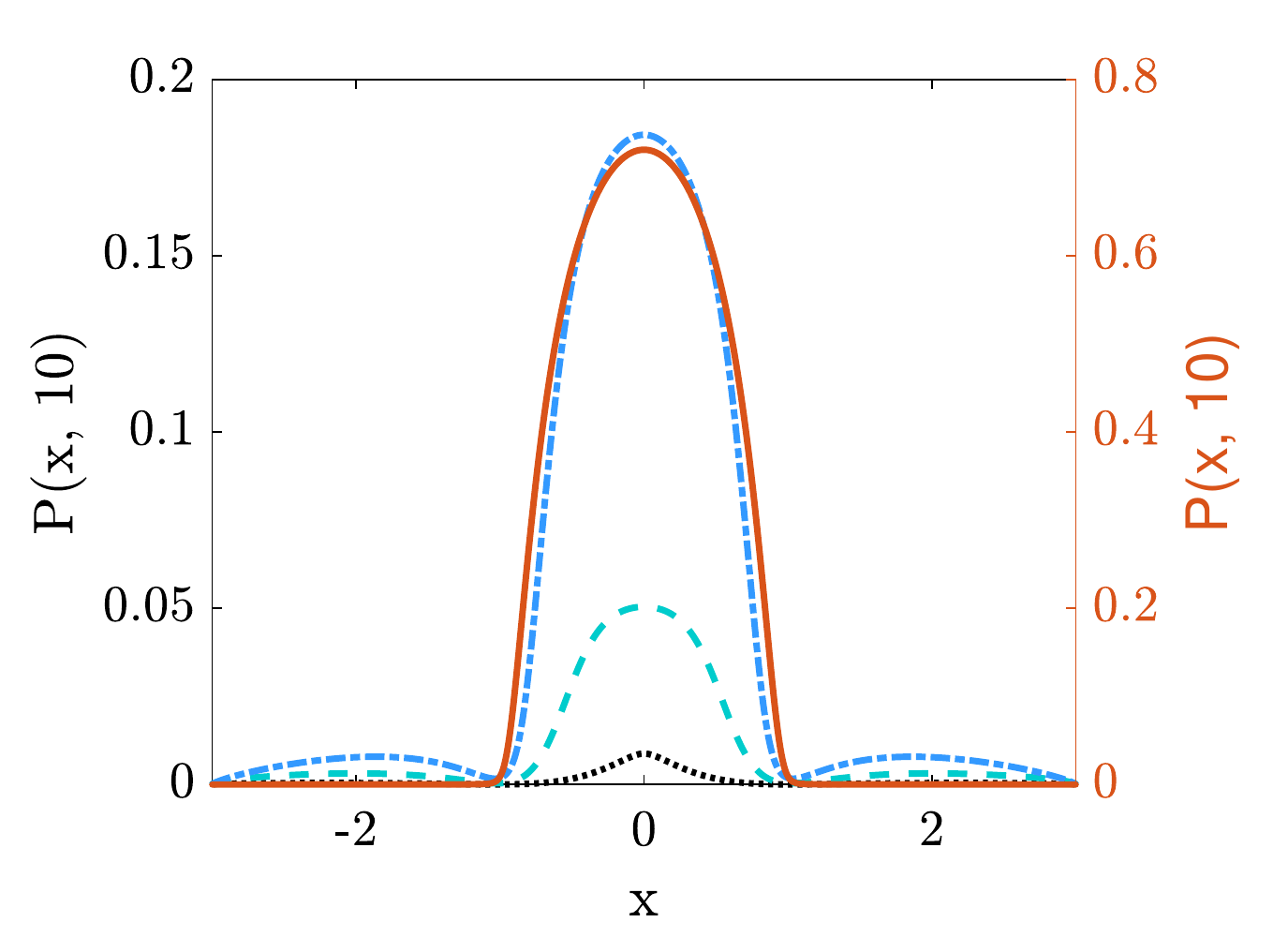}
\includegraphics[width=0.49\columnwidth]{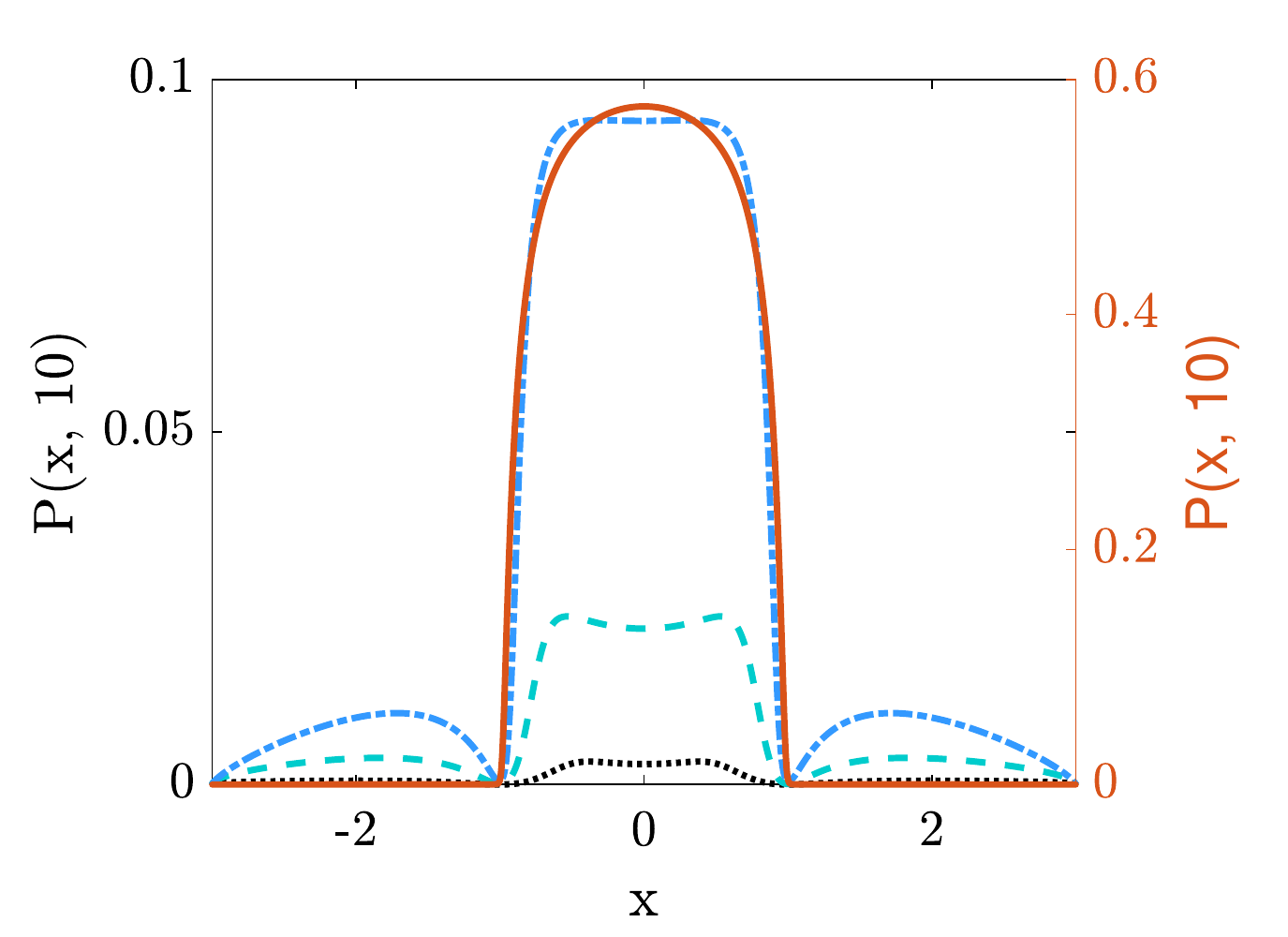}\\
\caption{The same as in Fig.~\ref{fig:fig1} for $h=10$.}
\label{fig:fig2}
\end{figure}

\begin{figure}[H]
\centering
\includegraphics[width=0.23\textwidth]{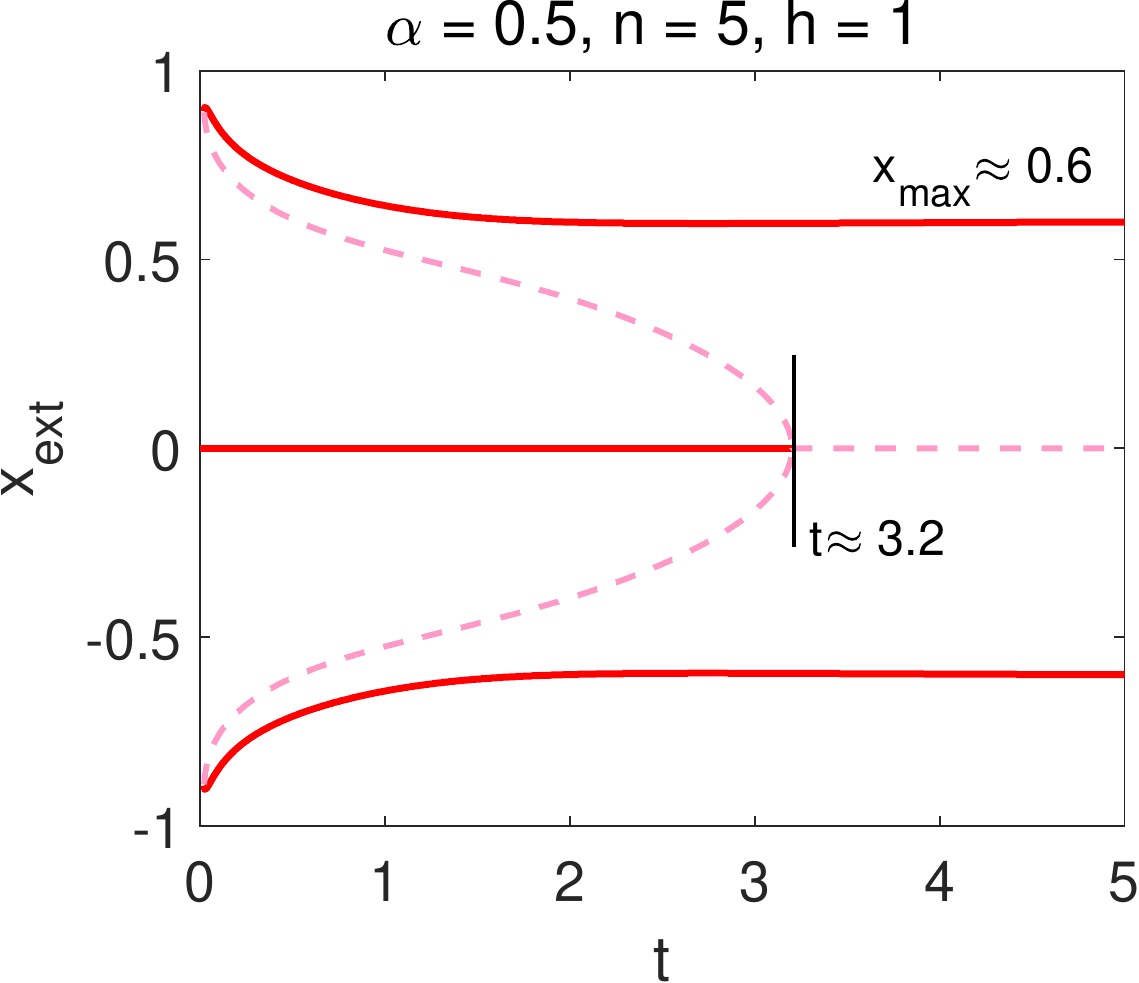}
\includegraphics[width=0.23\textwidth]{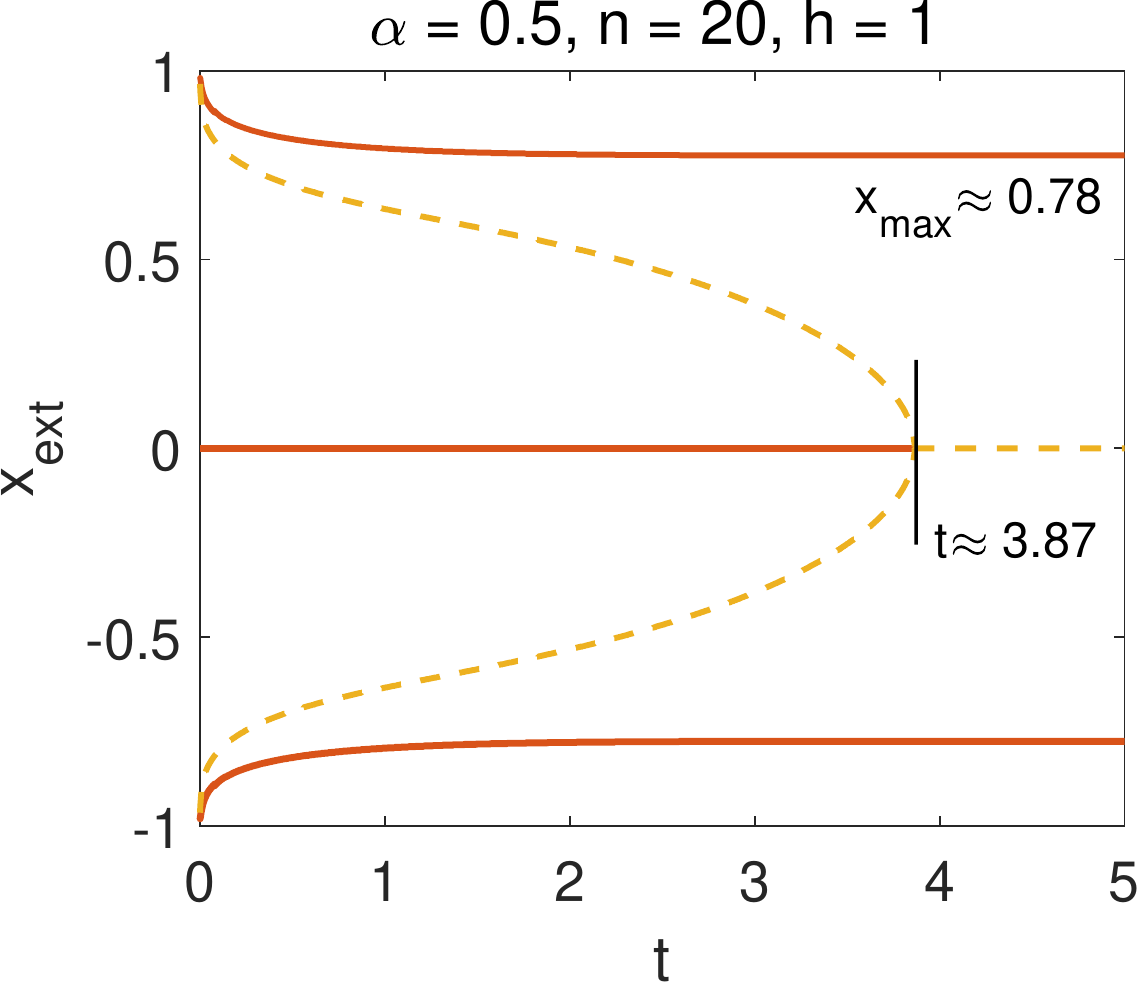}\\
\includegraphics[width=0.23\textwidth]{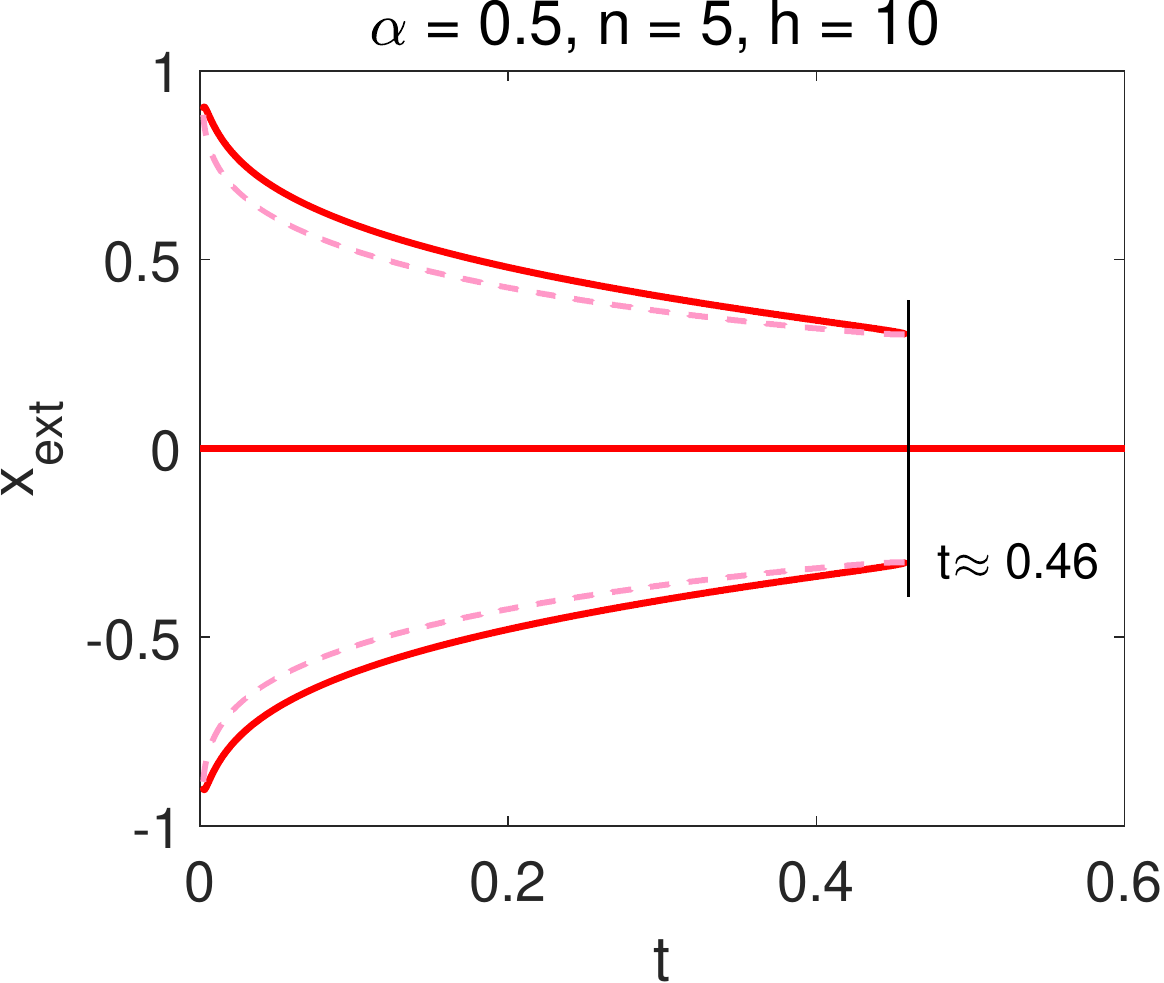}
\includegraphics[width=0.23\textwidth]{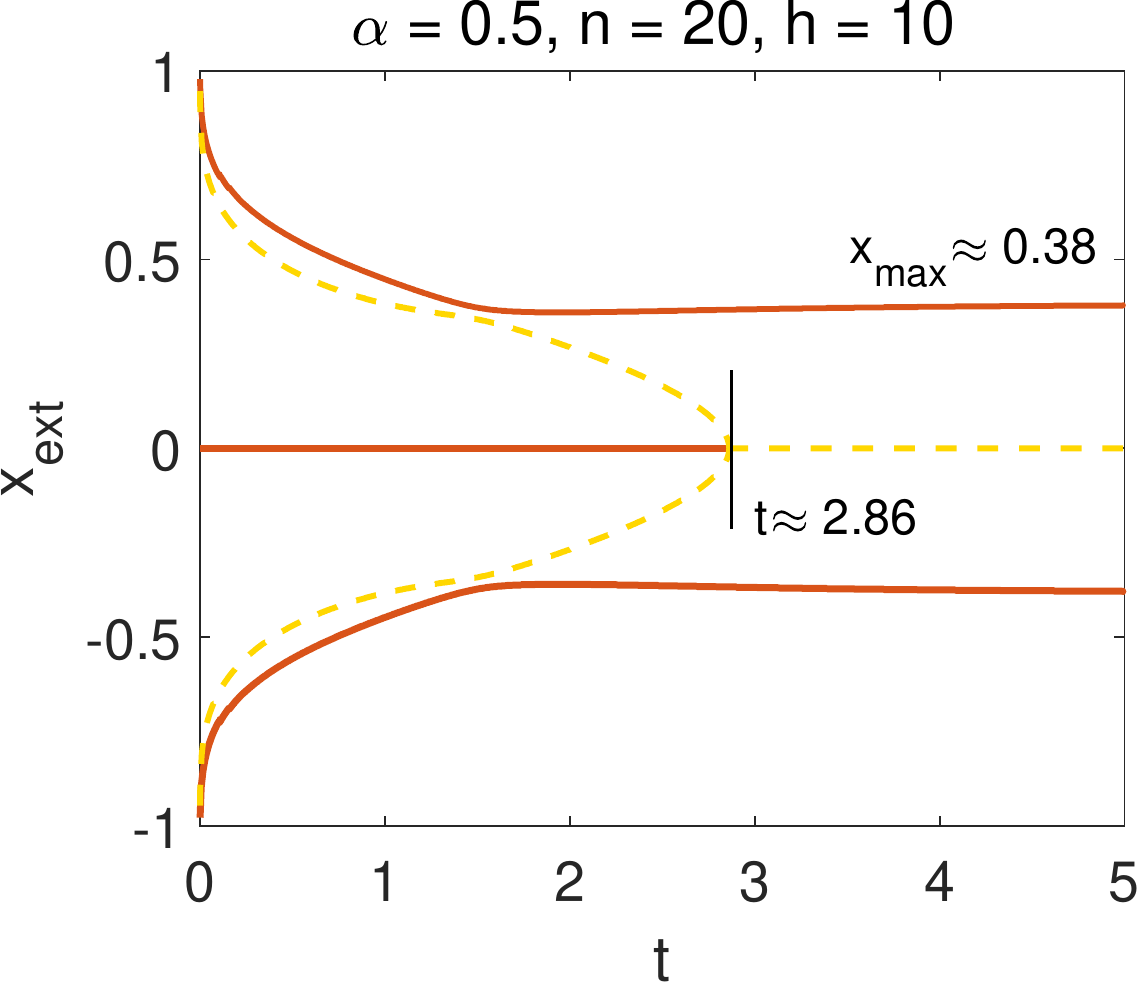}
\caption{The same as in Fig.~\ref{fig:bdiagram} for $\alpha=0.5$.}
\label{fig:fig3}
\end{figure}

\begin{figure}[H]
\centering
\includegraphics[width=0.23\textwidth]{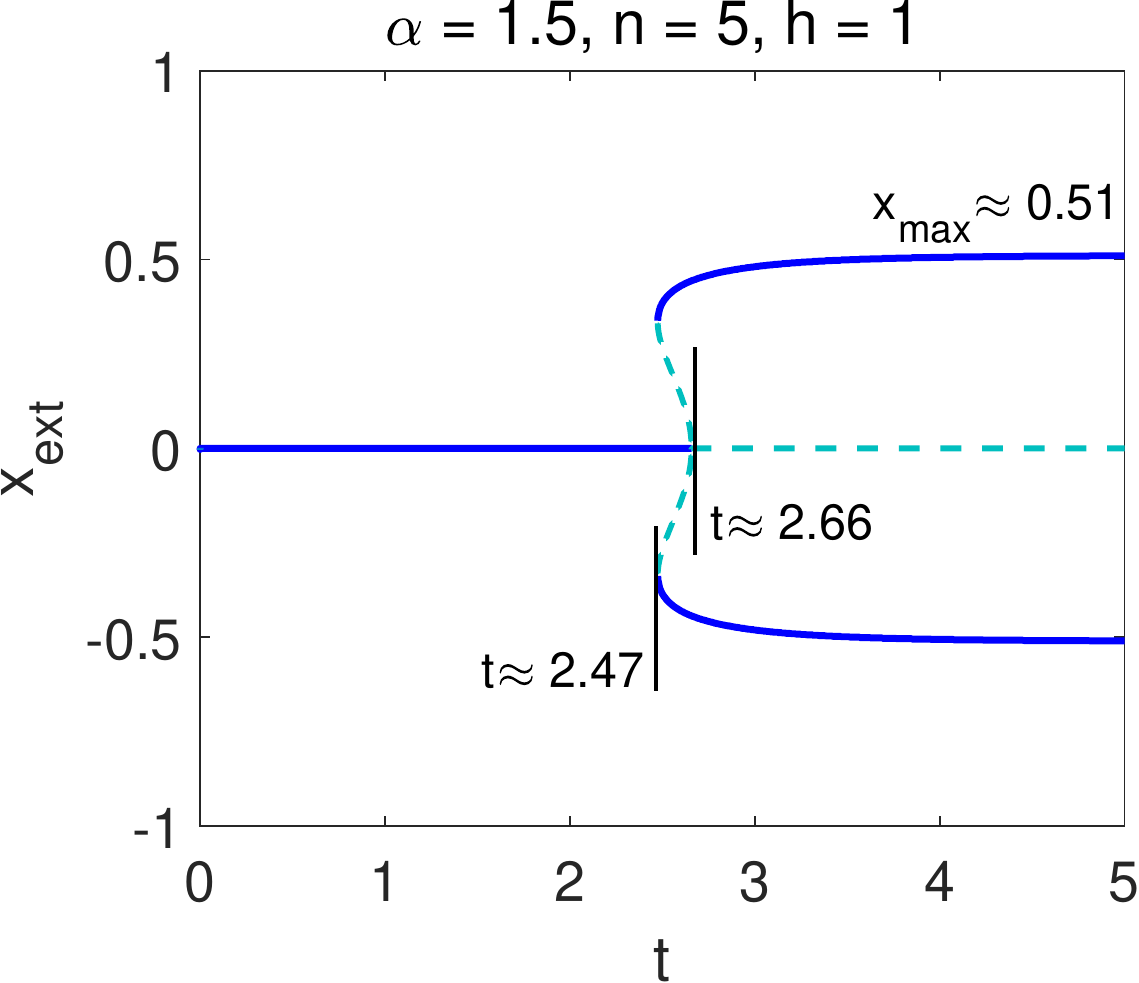}
\includegraphics[width=0.23\textwidth]{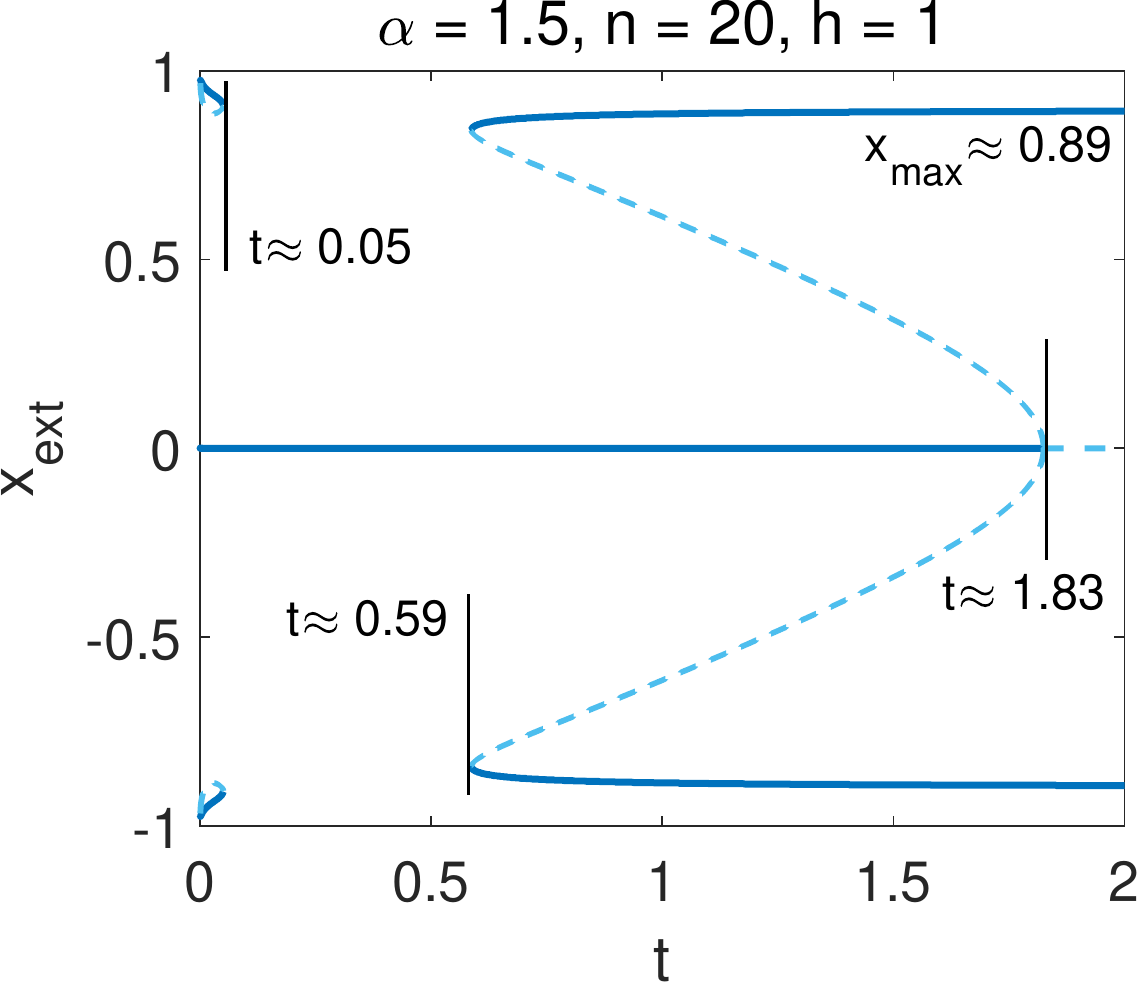}\\
\includegraphics[width=0.23\textwidth]{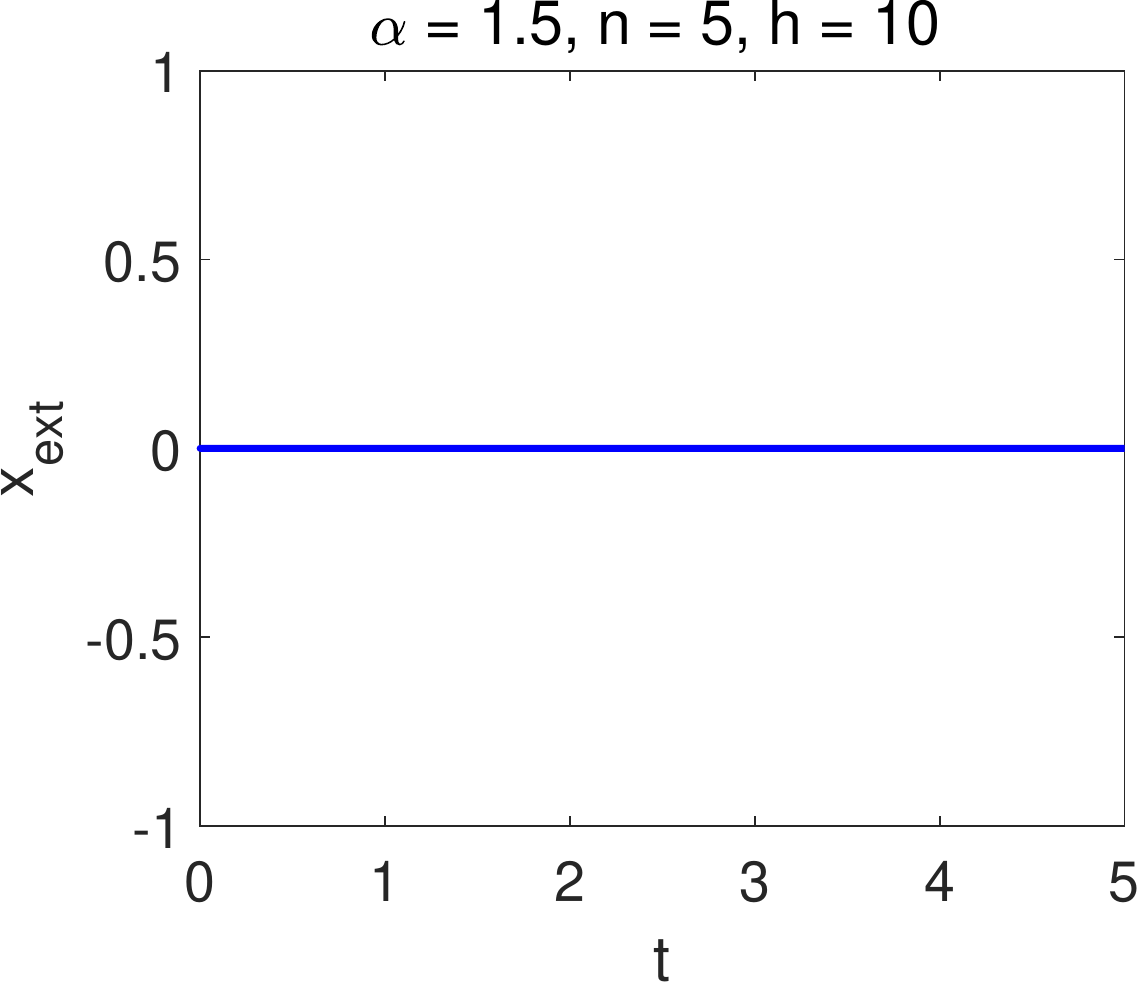}
\includegraphics[width=0.23\textwidth]{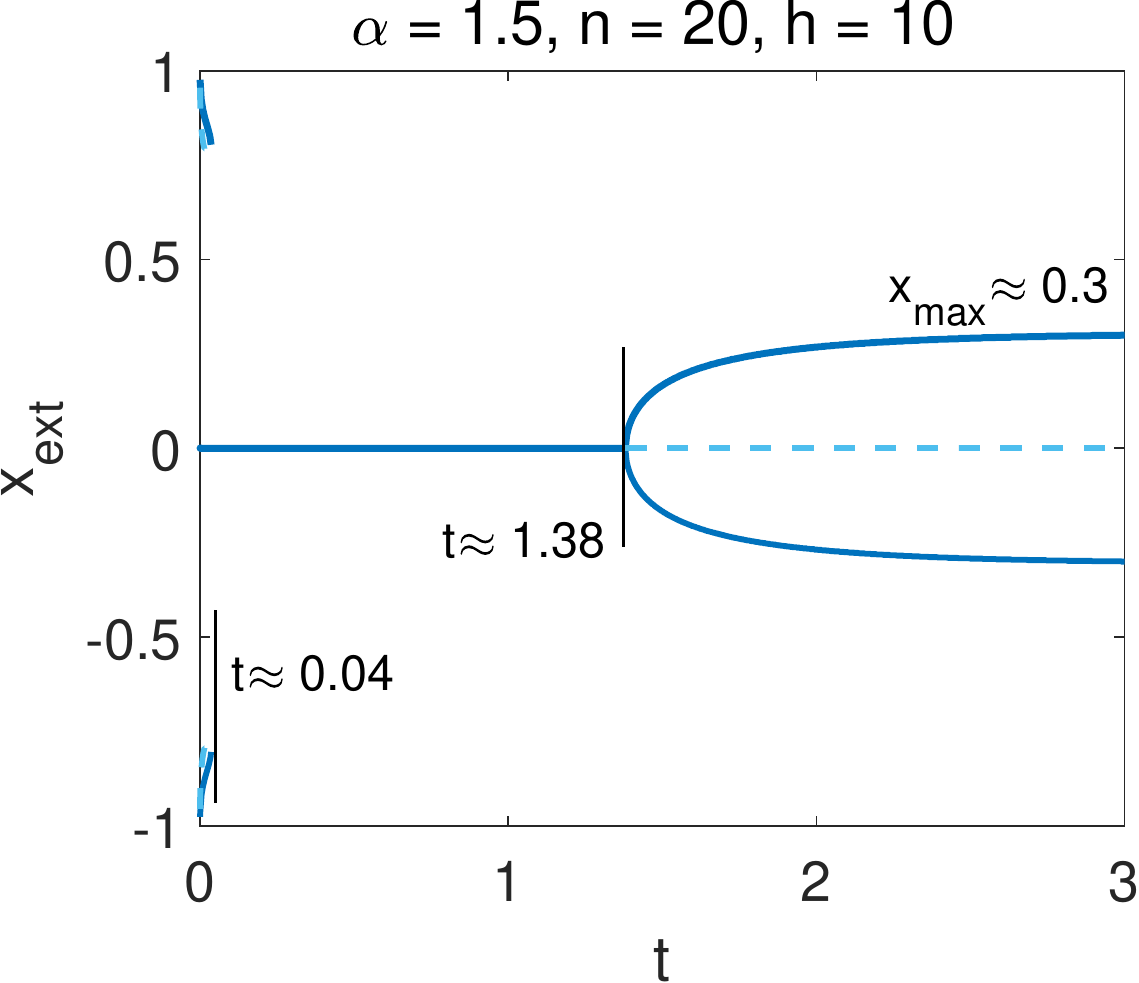}
\caption{The same as in Fig.~\ref{fig:bdiagram} for $\alpha=1.5$.}
\label{fig:fig5}
\end{figure}

While in the remainder of the text we concentrate on the guiding example of Cauchy ($\alpha=1$) white noise, in this section we present a comparison of results for the time dependent PDFs $P(x,t|x_0,0)$ with stable indices $\alpha=0.5$, $1$, and $1.5$ for different parameters $n$ and $h$, as depicted in Figs.~\ref{fig:fig1} and \ref{fig:fig2}.
\bdt{For clarity reasons, in the top panels of Figs.~\ref{fig:fig1} and \ref{fig:fig2}, we have restricted the range of the ordinate axis.
Otherwise, single peaks of $P(x,0.02)$ would hide details of the lower parts of the probability densities.}
Moreover, in Figs.~\ref{fig:fig3} and \ref{fig:fig5} we demonstrate the bifurcation diagrams showing crossovers between different shapes of the within the potential well PDFs with one, two and three maxima for the cases $\alpha=0.5$ and $1.5$, respectively.

In the left panels of Fig.~\ref{fig:fig1} we show results for the PDF $P(x,t|x_0=0,0)$ in the presence of the external \bdt{arctangent} potential~(\ref{eq:potential}) with $n=5$ and $h=1$ for different times.
As can be seen, for $\alpha=0.5$ the trimodal intrawell state with one main and two additional symmetric peaks emerge early on, while for $\alpha=1$ and $\alpha=1.5$ at short times there still exists the unimodal intrawell state.
In a more detailed analysis (not shown) we checked that this unimodal state turns into a trimodal state at times $t\approx1.1$ and $t\approx2.47$ (not shown), respectively, see also the top left panels in Figs.~\ref{fig:bdiagram} as well as \ref{fig:fig3} and \ref{fig:fig5}.
At intermediate times the trimodal state changes into a bimodal state at times $t\approx3.2$, $t\approx1.63$, and $t\approx 2.66$ for $\alpha=0.5$, $1$, and $1.5$, respectively, see also the top left panels in Figs.~\ref{fig:fig3} and \ref{fig:fig5}.

In the right panels of Fig.~\ref{fig:fig1} the PDF $P(x,t|x_0=0,0)$ is shown for the same fixed $h=1$ and the same times as in the left panels, however, for the larger value $n=20$.
As it is clear from the figure, with increasing steepness parameter $n$ the lifetime of the trimodal intrawell state for $\alpha=0.5$ increases, see also the top right panel of Fig.~\ref{fig:fig3}.
For $\alpha=1$ the trimodal intrawell state develops already at very short times and then changes into the bimodal intrawell state at $t\approx2.05$, see also the top right panel of Fig.~\ref{fig:bdiagram}.
Moreover, in the case of $\alpha=1.5$ we recognize a crossover from a trimodal to a unimodal state and back at times $t\approx0.05$ and $t\approx0.59$, respectively, while at $t\approx1.83$ the crossover to the bimodal state occurs, see also the top right panel of Fig.~\ref{fig:fig5}.

Similarly, in Fig.~\ref{fig:fig2} we demonstrate the time dependent PDFs with parameters $n=5$, $h=10$ (left panels), and $n=20$, $h=10$ (right panels) at different times. For $\alpha=0.5$ and $\alpha=1$ in the potential well on the left there are two additional symmetric peaks which disappear at short times $t\approx0.46$ and $t\approx0.09$, respectively, while for $\alpha=1.5$ we have a unimodal intrawell state at all times, see also the bottom left panels in Figs.~\ref{fig:fig3} and \ref{fig:fig5}.
In the right panels, the lifetime of the trimodal intrawell state decreases for $\alpha=0.5$, see the bottom right panel of Fig.~\ref{fig:fig3}. For $\alpha=1$ there is the crossover from a trimodal to a unimodal state and back at times $t\approx0.19$ and $t\approx1.14$, respectively, then at time $t\approx1.27$ a crossover to a bimodal state takes place, see the bottom right panel of Fig.~~\ref{fig:bdiagram}.
For $\alpha=1.5$ there occurs a crossover from a trimodal to a unimodal state at time $t\approx0.04$ and from a unimodal to a bimodal state at $t\approx1.38$, see the bottom right panel of Fig.~\ref{fig:fig5}.

\bdt{
Additionally, in Figs.~\ref{fig:fig1} and~\ref{fig:fig2}, we have depicted time the dependent densities for the Gaussian white noise driving.
The escape for $\alpha=2$ is significantly slower than for $\alpha<2$, consequently, under Gaussian white noise driving less particles become absorbed.
This increases the probability densities $P(x,t|x_0=0,0)$ with $\alpha=2$ compared to analogous densities with $\alpha<2$.
With increasing time this effect becomes amplified.
For clarity, therefore, for $t=10$ (bottom panels of Figs.~\ref{fig:fig1} and~\ref{fig:fig2}) values of $P(x,t|x_0=0,0)$ with $\alpha=2$ are plotted using right vertical axes.
For $\alpha=2$ the time dependent densities are  clearly unimodal.
Moreover, for $t \gg 1$, if the potential well is deep enough, e.g., $h=10$, the escape rate is so low that time dependent densities can be very well approximated by the quasi-stationary density given by Eq.~(\ref{eq:quastiStationaryStates}).
It is given by the renormalized Boltzmann--Gibbs state, i.e., $P(x,t|x_0=0,0)  \approx N S(t) \exp[-V(x)]$, see Fig.~\ref{fig:fig2}, where $S(t)$ is the fraction of particles (survival probability) which stay in the system and $N$ is the normalization factor $N=1/\int_{-3}^{3} \exp[-V(x)]dx$.
For $h=1$, such an approximation does not fully work.
It can indeed provide an approximation within the well part of $P(x,t)$, but it cannot reproduce nonzero probabilities outside the potential well, i.e., for $|x|>1$.
}

In Figs.~\ref{fig:fig3} and \ref{fig:fig5} we demonstrate the bifurcation diagrams for stability indices $\alpha=0.5$, $1$, and $1.5$, respectively. As can be seen, depending on the parameters $n$ and $h$, different crossovers between the unimodal, bimodal, and trimodal intrawell states occur.
For example, for the case $n=5$, $h=1$ for $\alpha=0.5$ we observe the crossover from trimodal to bimodal states at the bifurcation time $t\approx3.2$, see the top left panel of Fig.~\ref{fig:fig3}, while for $\alpha=1$ and $\alpha=1.5$ the crossover from unimodality to trimodality occurs at times $t\approx 1.1$ and $t\approx 2.47$, see the top left panel of Fig.~\ref{fig:bdiagram} as well as the top left panel of Fig.~\ref{fig:fig5}, respectively.
Moreover, a crossover from a trimodal to a bimodal state occurs at the bifurcation time $t\approx1.63$ for
$\alpha=1$ and at $t\approx2.66$ for $\alpha=1.5$.

For the stability index $\alpha=0.5$, on increasing parameter $n$ from 5 to 20 with the same $h=1$ the crossover time from the trimodal to the bimodal intrawell state increases from $t\approx3.2$ to $t\approx3.87$, see the top panels in Fig.~\ref{fig:fig3}. For $\alpha=1$ and $h=1$, with increasing $n$ from 5 to 20 the unimodality disappears, the lifetime of the trimodal state increases, and the bimodal state emerges at time $t \approx2.05$, see the top right panel of Fig.~\ref{fig:bdiagram}.
Moreover, for $\alpha=1.5$ in the cases $h=1$ and $n=20$, the crossover from the trimodal to the unimodal state occurs at the short time $t\approx0.05$, then the trimodal state emerges from the unimodal one at $t\approx0.59$, and finally a crossover from the trimodal to the bimodal state at $t \approx1.83$ is observed, see the right panel of Fig.~\ref{fig:fig5}.

With the same $n=5$, by increasing $h$ from 1 to 10 for the stability indices $\alpha=0.5$ and $\alpha=1$, there exist crossovers from trimodal to unimodal intrawell states at the bifurcation times $t\approx0.46$ and $t\approx 0.09$, respectively, see the bottom left panels in Figs.~\ref{fig:bdiagram} and~\ref{fig:fig3}.
At the same time, for $\alpha=1.5$ with $n=5$ and $h=10$, the PDF is always unimodal, see the bottom left panel of Fig.~\ref{fig:fig5}.

Finally, we notice that for $\alpha=0.5$ with $n=20$, by increasing the parameter $h$ from 1 to 10 the lifetime of the trimodal intrawell state decreases from $t\approx3.87$ to $t\approx2.86$, see the right panels in Fig.~\ref{fig:fig3}. For $\alpha=1$ with $n=20$ and $h=10$, the crossovers occur from a trimodal to a unimodal state at $t\approx0.19$, then back from unimodal to trimodal at $t\approx1.14$, and finally from trimodal to bimodal at $t\approx1.27$, see the bottom right panel of Fig.~\ref{fig:bdiagram}.
Moreover, for $\alpha=1.5$ with $n=20$ and $h=10$, the trimodal-unimodal crossover appears at $t\approx0.04$, and the unimodal-bimodal crossover at $t \approx1.38$, see the bottom right panel of Fig.~\ref{fig:fig5}.

%%%%%%%%%%%%%%%%%%%%%%%%%%%%%%%%%%%%%%%%%%%%%%%%%%%%%%%%%%
% \clearpage
\section{Conclusions \label{sec:summary}}

Noise is a well established and frequently used concept in statistical physics.
It is employed to approximate complicated, irregular collisions of a test particle with its environment.
Moreover, L\'evy flights are applied in various biological applications including search strategies \cite{sims2011,palyulin2014levy,viswanathan2011physics}.
In the context of the current research there are two very related problems induced by L\'evy noise: the escape from potential wells and the emergence of stationary states.

Within the current study we examined the problem of L\'evy noise-driven escape of a particle from a box-like potential well.
Moreover, we assumed that the whole domain of motion is restricted by two absorbing boundaries.
We focused on the qualification of the time-dependent densities, the last hitting point densities and the first passage problems.

Time dependent densities  are constituted of two parts.
The part located within the potential well is similar to the stationary densities recorded in single-well potentials.
Analogously, like in single-well potentials, the time dependent densities can be transiently trimodal.
The outer part of the time dependent densities is determined by the shape of the $\alpha$-stable densities
and diffusion along the flat part of the potential profile.
Due to the presence of absorbing boundaries the probability of finding a particle in the domain of motion
decays exponentially over time.

The examination of the last hitting point densities reveals possible escape scenarios.
A particle can escape from the system via a single long jump or in a sequence of short jumps.
Long jumps relate the last hitting point density to the time dependent densities, while short jumps produce peaks in the vicinity of the absorbing boundaries.

Finally, the examination of the first passage problem confirms that for high potential barriers, the escape time is determined by the barrier width.
Moreover, the substantial contribution to the first passage time is produced by a part of the motion which explores flat parts of the potential profile.

%%%%%%%%%%%%%%%%%%%%%%%%%%%%%%%%%%%%%%%%%%

\begin{acknowledgments}
KC was supported by the Faculty of Physics, Astronomy and Applied Computer Science under the DSC 2019-N17/MNS/000013 scheme.
This research was supported in part by PLGrid Infrastructure. Computer simulations were performed at Shahid Beheshti University (Tehran, Iran) and at the Academic Computer Center Cyfronet, AGH University of Science and Technology (Krak\'ow, Poland).
AC and RM acknowledge support from German Science Foundation (DFG, grant no. ME 1535/7-1).
RM also thanks the Foundation for Polish Science (Fundacja na rzecz Nauki Polskiej, FNP) for support within an Alexander von Humboldt Polish Honorary Research Scholarship.
\end{acknowledgments}

%%%%%%%%%%%%%%%%%%%%%%%%%%%%%%%%%%%%%%%%%%

\section*{Data availability}
The data that support the findings of this study are available from the
corresponding author (KC) upon reasonable request.

%\section*{References}
% \bibliographystyle{iopart-num}
%\bibliographystyle{apsrev}
%\bibliography{core-bibliography}

\def\url#1{}

\end{document}